\documentclass[reprint,pra,aps,nofootinbib,longbibliography,superscriptaddress]{revtex4-2}

\usepackage{amsmath,amssymb,amstext}
\usepackage[pdftex]{graphicx}
\usepackage{tabularx}
\usepackage{bm}
\usepackage{mathrsfs}
\usepackage{fancyhdr}
\usepackage{booktabs}
\usepackage{longtable}
\usepackage{multirow}
\usepackage{float}
\usepackage{outlines}

\usepackage[pdftex,pagebackref=false]{hyperref}
\hypersetup{
    unicode=false,          
    pdffitwindow=false,     
    pdfstartview={FitH},    
    pdfnewwindow=true,      
    colorlinks=true,        
    linkcolor=blue,         
    citecolor=green,        
    filecolor=magenta,      
    urlcolor=cyan           
}

\usepackage{braket}
\usepackage{amsthm}
\usepackage{bbold}
\usepackage{comment}

\allowdisplaybreaks

\newcommand{\mf}{\mathsf}

\newtheorem*{proposition*}{Proposition}

\begin{document}

\title{Harvesting Contextuality from the Vacuum}

\author{Philip A. LeMaitre}
\email{Philip.Lemaitre@uibk.ac.at}
\affiliation{University of Innsbruck, Institute for Theoretical Physics, Technikerstrasse 21a, A-6020 Innsbruck, Austria}

\date{\today}

\begin{abstract}
Quantum contextuality is the notion that certain measurement scenarios do not admit a global description of their statistics and has been implicated as the source of quantum advantage in a number of quantum information protocols. It has been shown that contextuality generalizes the concepts of non-local entanglement and magic, and is an equivalent notion of non-classicality to Wigner negativity. In this paper, the protocol of contextuality harvesting is introduced and it is shown that Unruh-DeWitt models are capable of harvesting quantum contextuality from the vacuum of a massless scalar quantum field. In particular, it is shown that gapless systems can be made to harvest contextuality given a suitable choice of measurements. The harvested contextuality is also seen to behave similarly to harvested magic and can be larger in magnitude for specific parameter regimes. An Unruh-DeWitt qubit-qutrit system is also investigated, where it is shown that certain tradeoffs exist between the harvested contextuality of the qutrit and the harvested entanglement between the systems, and that there are regimes where the two resources can both be present. Some of the tools of contextuality, namely the contextual fraction, are also imported and used as general measures for any form of harvested contextuality, including non-local entanglement and magic. Additionally, new criteria for genuine harvesting are put forward that also apply to individual systems, revealing new permissible harvesting parameter regimes.
\end{abstract}

\maketitle

\section{Introduction} \label{Section:Introduction}

It is by now well-documented in the relativistic quantum information (RQI) literature that relativistic quantum fields are full of information-theoretic resources~\cite{Halvorson_2007,Haag,Butterﬁeld_Halvorson,hall2013quantum}, such as entanglement and magic, that one can extract for use in various quantum information protocols such as entanglement harvesting \cite{Valentini1991,Reznik2003,Pozas-Kerstjens_Martín-Martínez_2016,Maeso-García_Polo-Gómez_Martín-Martínez_2022,Perche_Polo-Gómez_Torres_Martín-Martínez_2023}, magic harvesting~\cite{Nyström_Pranzini_Keski-Vakkuri_2024}, quantum energy teleportation \cite{hotta2008quantum,HottaDistance,HottaEntanglement,teleportExperiment}, and quantum field-mediated communication \cite{Jonsson_Ried_Martin-Martinez_Kempf_2018,Simidzija_Ahmadzadegan_Kempf_Martín-Martínez_2020}, or to power a relativistic quantum computer \cite{LeMaitre_Perche_Krumm_Briegel_2024,Friis_Bruschi_Louko_Fuentes_2012,AsplingLawler2024,Bruschi_Dragan_Lee_Fuentes_Louko_2013,Layden_Martin-Martinez_Kempf_2016,Martin-Martinez_Aasen_Kempf_2013}. While these resources are considered useful under their respective resource theories, one can't help but ask whether there are more general and as yet undiscovered resources lurking inside quantum fields. As it turns out, there is a non-classical resource which encompasses and moves beyond most if not all of the previously encountered resources: quantum contextuality ~\cite{Kitajima_2006,Kitajima_2017,Schlichtholz_Mandarino_Żukowski_2022}.

The statement of contextuality is essentially the inability to assign a joint probability distribution to a whole family of measurements that contain subsets of compatible measurements, i.e. one cannot consistently marginalize a joint probability distribution over all measurements to yield the distributions for all of the individual subsets of measurements in the family \cite{Abramsky_Barbosa_Mansfield_2017,abramsky2011sheaf}. The notion of non-locality then falls out of this description when the system and measurements are considered to be composite (and typically spatially separated); it is this form of contextuality in quantum fields which has been extensively studied~\cite{Summers_Werner,Summers_Werner_1985,Summers_Werner_1995, Summers_Werner_1987a,Summers_Werner_1987b,Summers_Werner_1987c,Schlichtholz_Markiewicz_2024,Benatti_Floreanini_Narnhofer_2025,Butterﬁeld_Halvorson,Clifton_Halvorson_2001,DeFabritiis_Roditi_Sorella_2023,DeFabritiis_Guedes_Guimaraes_Roditi_Sorella_2024,Grossi_Barata_2024,Guedes_Guimaraes_Roditi_Sorella_2024,Valente_2014,Rédei_1989,Rédei_1991}. Contextuality has been experimentally verified in numerous settings \cite{Ru_Tang_Wang_Wang_Zhang_li_2022,Hu_Liu_Chen_Guo_Wu_Huang_Li_Guo_2018,Xue_Xiao_Ruffolo_Mazzari_Temistocles_Cunha_Rabelo_2023,Mazurek_Pusey_Kunjwal_Resch_Spekkens_2016} and has been identified as the source for quantum advantage in many quantum information protocols \cite{Gupta_Saha_Xu_Cabello_Majumdar_2023,Shahandeh_2021,Emeriau_Howard_Mansfield_2022,Anschuetz_Hu_Huang_Gao_2023,Schmid_Spekkens_2018}, in addition to those of non-local entanglement ~\cite{abramsky2011sheaf}. Contextuality and Wigner negativity were also shown to be equivalent notions of non-classicality for continuous-variable quantum systems~\cite{Haferkamp_Bermejo-Vega_2021,Booth_Chabaud_Emeriau_2022}, qudits in odd-prime-dimensional Hilbert spaces\footnote{Wigner negativity in the standard discrete Wigner function formalism ~\cite{Gross_2006,Cormick_Galvao_Gottesman_Paz_Pittenger_2006,Galvão_2005} is not strictly equivalent to contextuality for qubits, but is fine for the extension used in Reference ~\cite{Kocia_Love_2017}.}~\cite{Delfosse_Okay_Bermejo-Vega_Browne_Raussendorf_2017,Raussendorf_Browne_Delfosse_Okay_Bermejo-Vega_2017,Schmid_Du_Selby_Pusey_2022}, and in general when adapting Wigner negativity to the ontological models framework and comparing it with generalized contextuality ~\cite{Spekkens_2008,Ferrie_Emerson_2009}. The discrete Wigner function formalism ~\cite{Gross_2006,Cormick_Galvao_Gottesman_Paz_Pittenger_2006,Galvão_2005} was also used to link Wigner negativity to magic, and to create the magic measure known as the mana ~\cite{Veitch_Ferrie_Gross_Emerson_2012,Veitch_Mousavian_Gottesman_Emerson_2014}. Then it was shown in Reference \cite{Howard_Wallman_Veitch_Emerson_2014} that magic states for qudits in Hilbert spaces of odd prime dimension are necessarily contextual with respect to stabilizer measurements.


The presence of contextuality also serves as a marker of non-classicality that is explicitly clear in the resource theory of contextuality~\cite{Amaral_2019}: one always defines the classical state space boundary through their choice of measurement settings. This gives a much more straightforward methodology to interpret the utility of any acquired resources compared to, say, entanglement. While it is still useful to study the role of entanglement and other resources in RQI protocols, it is important to delineate the boundary between what states one considers to be classical versus non-classical, if the intention is to discuss genuine quantumness or quantum advantage in RQI.


With all of the demonstrated advantages due to contextuality, the main question that will be tackled in this paper is whether a system in an initially non-contextual state can evolve to a contextual state after interacting with a quantum field\footnote{Close to the completion of this paper, the author became aware of similar, independent work by C. Lima, M. Rosa Preciado-Rivas, and S. Srivastava ~\cite{Lima_Preciado-Rivas_Srivastava_2025}}. 
Additionally, how does contextuality behave as a resource when considered in its own right and also when compared to, say, magic? Furthermore, how do the various types of contextuality (magic, local contextuality, non-locality, etc.) and entanglement interact with each other? Can more than one resource be present in the system at a time? Some works in the non-relativistic setting \cite{Xue_Xiao_Ruffolo_Mazzari_Temistocles_Cunha_Rabelo_2023,Porto_Ruffolo_Rabelo_Cunha_Kurzynski_2024,Hu_Liu_Chen_Guo_Wu_Huang_Li_Guo_2018} have shown that there do indeed exist settings where entanglement and local contextuality can coexist and influence the amount of each other in the system, with tradeoff relations bounding the total amount of these resources that can be present at a time \cite{Porto_Ruffolo_Rabelo_Cunha_Kurzynski_2024}. Meanwhile, other works \cite{Cabello_2021,Plávala_Gühne_2024,Wright_Kunjwal_2023,Wright_Farkas_2023} have elucidated the role that contextuality plays in witnessing non-locality and what non-locality implies about contextuality. Thus, it would be interesting to investigate how these results manifest themselves in the relativistic setting and what the exact mechanisms which give rise to them would look like. 


To answer these questions, some measurement scenarios and measures of contextuality will be introduced and then applied to the setting of the Unruh-DeWitt (UDW) model \cite{PhysRevD.14.870,Hawking:1979ig} for a single qutrit and qubit-qutrit systems; single qutrits because they are the simplest systems to exhibit contextuality, and a qubit-qutrit system because it is the simplest system to witness both qutrit contextuality and entanglement. This manuscript is then organized as follows. A brief introduction to quantum contextuality is given in Section \ref{Section:Contextuality} along with the measurement scenarios and measures of contextuality that are considered. Next, the UDW model is introduced in Section \ref{Section:UdW} and the specific setup considered is detailed. Then in Section \ref{Section:ContextHarvest} the results of the contextuality harvesting experiments are introduced, and finally in Section \ref{Section:Conclusions} the implications of the results are discussed and some future research avenues are outlined.

\section{Quantum Contextuality} \label{Section:Contextuality}

The basics of quantum contextuality will be introduced here as they are essential to understanding the results presented later in Section \ref{Section:ContextHarvest}. The presence of quantum contextuality in a system is often taken as the definition of what it means for that system to be considered non-classical. It is consistently the phenomenon that resists an explanation in terms of local hidden variables, which almost all other phenomena in quantum theory can be described by ~\cite{Spekkens_2007,Catani_Leifer_Scala_Schmid_Spekkens_2022,Catani_Leifer_Schmid_Spekkens_2023}. There are many approaches to contextuality \cite{amaral2018graph,Spekkens_2005,abramsky2011sheaf,Okay_Kharoof_Ipek_2023,Dzhafarov_Kujala_Cervantes_2015} which have expanded it to be a generic feature of generalized probabilistic theories, of which quantum theory is a special case \cite{Schmid_Selby_Rossi_Baldijão_Sainz_2024}. 
In anticipation of using the contextual fraction \cite{Abramsky_Barbosa_Mansfield_2017} to measure the level of contextuality, the sheaf-theoretic approach to contextuality will be adopted, and so a few objects from the sheaf-theoretic framework of contextuality will need to be introduced before contextuality can be understood. To this end, the presentation will closely follow that of reference \cite{Abramsky_Barbosa_Mansfield_2017}. 

The central objects of study in the sheaf-theoretic picture are called empirical models, which are probability data tables for the joint outcomes of sets of compatible measurements corresponding to a measurement scenario $M \equiv (X, \mathcal C, O)$; a tuple consisting of a finite set of measurements $X$, a finite set of outcome values for each measurement $O$ (sometimes denoted as $O^X$), and a set of subsets of X denoted as $\mathcal C$ . Each $C \in \mathcal C$ is called a measurement context and represents a collection of measurement operators that can be performed simultaneously. As an example, consider the classic Klyachko-Can-Binicioğlu-Shumovsky (KCBS) scenario \cite{amaral2018graph,klyachko2008simple} for a single party who can choose to measure one of $\{m_i\}_{i=0}^4$, obtaining one of two possible outcomes: 
\begin{align}\label{Equation:BellScen}
&X = \{m_0, m_1, m_2, m_3, m_4\}\ , \ O = \{-1, 1\}\ ,\nonumber \\ \mathcal C = \{&\{m_0,m_1\},\{m_1,m_2\},\{m_2,m_3\},\{m_3,m_4\}, \{m_4, m_0\}\}
\end{align}
\begin{table}
    \centering
    \begin{tabular}{|c|c|l|l|l|} \hline  
 & $-1, -1$& $-1, 1$& $1, -1$&$1, 1$\\ \hline  
         $m_0, m_1$&   0& 1/9& 2/3&2/9\\ \hline  
         $m_1, m_2$& 
     0& 2/3& 1/3&0\\ \hline  
 $m_2, m_3$& 0& 1/3& 1/3&1/3\\ \hline  
 $m_3, m_4$& 0& 1/3& 2/3&0\\ \hline
 $m_4, m_0$& 0& 2/3& 1/9&2/9\\\hline \end{tabular}
    \caption{An empirical model corresponding to the KCBS scenario Eq. \eqref{Equation:BellScen}.}
    \label{Table:222Bell}
\end{table}
The empirical model for this description can then be constructed either from theoretical predictions or through multiple runs of an experiment, resulting in a probability table such as Table \ref{Table:222Bell}. 

The probability distributions making up an empirical model are the joint distributions $e_C$ for each context $C \in \mathcal C$ over the joint outcome space denoted by $O^C$, with the requirement that the marginals of these distributions, denoted $e_C|_U$ with $U \subseteq C$, agree whenever contexts overlap: $e_C|_{C \cap C'} = e_{C'}|_{C\cap C'} \ \forall C, C' \in \mathcal C$; a generalization of the no-signalling condition. Contextuality within an empirical model is then the statement that no global distribution $d$ on $O^X$ exists such that its marginals give the empirical model, i.e. $d|_C = e_C \ \forall C \in \mathcal C$. Equivalently, contextual empirical models are those which have no realization by factorizable hidden variable models, which for Bell-type measurement scenarios, means that contextuality reduces to the usual notion of non-locality that characterizes some forms of entanglement. Note also that standard contextuality can only be observed in systems with Hilbert space dimension greater than 3 \cite{peres1997quantum}.

For the most part, the contextuality of a system in state $\hat{\rho}$ with respect to scenario $M$ is assessed in a binary fashion through the calculation of what is known as a non-contextuality inequality:
\begin{align} \label{Equation:NonContextInEq}
S_{\mathcal C} = \sum_{i_C\in i_{\mathcal C}}h_{i_C}\text{Tr}\left(\hat{\rho}\hat{M}_{i_C}\right) \leq \gamma
\end{align}
where $i_C$ denotes the set of indices corresponding to the context $C \in \mathcal C$,  $h_{i_C} \in \{\pm 1\}$ are constants, $\hat{M}_{i_C}$ is the (tensor) product of operators indexed by the context $C$, and $\gamma$ is an upper bound that marks the boundary of the space of non-contextual states \cite{amaral2018graph}. In the setting of empirical models, the combination of eigenprojectors corresponding to the operators in each summand of Eq. \eqref{Equation:NonContextInEq} define one row of the associated empirical model. The compatibility relations between the $\hat{M}_{i_C}$ making up the scenario $M$ can be given a graph structure called a compatibility graph, whose maximum independence number - the size of the largest subgraph comprised of vertices that have no edges in common - is the constant $\gamma$ \cite{amaral2018graph}. If $\gamma$ in Eq. \eqref{Equation:NonContextInEq}  can be exceeded using $\hat{\rho}$, then the system is said to be contextual, otherwise it is non-contextual. Eq. \eqref{Equation:NonContextInEq} can be split into two different settings: state-dependent and state-independent non-contextuality inequalities. In both settings, the scenario $M$ shapes the boundary for the states $\hat{\rho}$ of a given Hilbert space dimension that can violate Eq. \eqref{Equation:NonContextInEq} , and so the difference lies in how large this boundary is. For the state-dependent case, the boundary is such that some states can violate Eq. \eqref{Equation:NonContextInEq}
while others cannot. In the state-independent case, as the name implies, there is no boundary and therefore all states can violate Eq. \eqref{Equation:NonContextInEq} regardless of their structure. 

Since the goal will be to harvest contextuality from the state of a quantum field, only the state-dependent scenarios will be of interest to us. Further anticipating that the effect of contextuality harvesting will be quite small, as is the case in entanglement harvesting \cite{Perche_2024,Pozas-Kerstjens_Martín-Martínez_2015,Pozas-Kerstjens_Martín-Martínez_2016}, it would be prudent to define a continuous measure of contextuality so the amount of contextuality harvested can be quantified in a non-binary way. To this end, the contextual fraction is adopted \cite{Abramsky_Barbosa_Mansfield_2017}.  

The contextual fraction is motivated by the fact that an empirical model can be decomposed as a convex sum of two different empirical models on the same scenario: $e\lambda +e'(1-\lambda)$, where $\lambda \in [0, 1]$; this can be understood by taking the convex sum of probability distributions at each context, which preserves the compatibility relations. If one of these empirical models is now viewed as a non-contextual model, then the constant $\lambda$ can be interpreted as giving the fraction of the original model that is explained by a non-contextual model, which provides a continuous measure as opposed to the binary one in Eq. \eqref{Equation:NonContextInEq}. Instead of asking for a probability distribution on global assignments that marginalizes to the empirical distributions at each context, only a subprobability distribution $ b$ on global assignments $O^X$ that marginalizes at each context to a subdistribution of the empirical data is asked for, thus explaining a fraction of the events, i.e. $b|_C \leq e_C \ \forall C \in \mathcal C$. When the maximum possible value of $\lambda$ is used in the decomposition, this is called the non-contextual fraction of $e$ and denoted $NCF(e)$; it is related to the contextual fraction by $CF(e) = 1-NCF(e)$. The other empirical model in the decomposition will then necessarily be a strongly contextual one \cite{Abramsky_Barbosa_Mansfield_2017}, which means any empirical model admits the decomposition 
\begin{align}
e = CF(e)e^{SC}+NCF(e)e^{NC}
\end{align}
where $e^{SC}$ denotes the strongly contextual empirical model and $e^{NC}$ the non-contextual empirical model (Note that $e^{NC}$ and $e^{SC}$ are not necessarily unique). 

The task of finding a consistent probability subdistribution with maximum weight for a given empirical model can be formulated as a linear programming problem. Define a measurement scenario $M \equiv (X, \mathcal C, O)$ and let $n = |O^X|$ be the number of global assignments, and $m = \sum_{C \in \mathcal C} |O^C|$ be the total number of local assignments for all contexts. Then define an $m\times n$ incidence matrix $\bm M$ that records the restriction relation between global $g$ and local $(C, s)$ assignments:
\begin{align}
\bm M[(C, s), g] = \begin{cases}
    1 & \text{if}\ g|_C = s \\
    0 & \text{else}
\end{cases}
\end{align}
The empirical model $e$ can be vectorized into a vector $\bm v^e \in \mathbb R^m$, where the component $\bm v^e[(C, s)] = e_C(s)$, i.e. the model probability for assignment $s$ in context $C$ (an entry in Table \ref{Table:222Bell}). Each column of $\bm M[-,g]$ can then be seen to represent one of the (non-contextual) deterministic models obtained from global assignments $g \in O^X$. A global probability distribution can also be represented as a non-negative normalized vector $\bm d \in \mathbb R^n$, so that the corresponding non-contextual model can be represented by the vector $\bm M \bm d$. Then a model $e$ is considered non-contextual if and only if there exists a non-negative $\bm d \in \mathbb R^n$ such that $\bm M \bm d \leq \bm v^e$. A global subprobability distribution is also represented by a non-negative vector $\bm b \in \mathbb R^n$, with its weight being given by the dot product $\bm 1 \cdot \bm b = \sum_i^n b_i$. The following linear program thus calculates the non-contextual fraction of an empirical model $e$, with $NCF(e)=\bm 1\cdot \bm b^{*}$, where $\bm b^{*}$ is an optimal solution:
\begin{align}
& \text{Find} & \bm b \in \mathbb R^n \nonumber \\
& \text{maximizing} & \bm 1 \cdot \bm b \nonumber\\
& \text{subject to} & \bm M \bm b \leq \bm v^e \nonumber \\
& \text{and} & \bm b \geq \bm 0
\end{align}
As shown in reference \cite{Abramsky_Barbosa_Mansfield_2017}, the contextual fraction can be related to non-contextuality inequalities as the normalized violation by a given empirical model of a non-contextuality inequality. The authors also prove that the contextual fraction is a resource monotone under operations that preserve the non-contextual set and that it shares many of the same properties that standard entanglement measures have \cite{Plenio_Virmani_2006}: faithfulness, convexity, and continuity.


In this article, a variant of the 5-measurement pentagram scenario \cite{Badziag_Bengtsson_Cabello_Granstrom_Larsson_2011} for a qutrit system will be considered due to its simplicity, flexibility with what states can be labelled as classical, and that it is somewhat well-studied ~\cite{Badziag_Bengtsson_Cabello_Granstrom_Larsson_2011}. It is characterized by 5 dichotomic measurement operators $\{\hat{B}_i|\hat{B}_i=\openone-2|v_i\rangle\langle v_i|\}_{i=0}^4$, each with outcome $\pm 1$,  where the operators satisfy $[\hat{B}_i, \hat{B}_j] = 0$ for the index pairs $\{(0,2), (0,3), (1,3), (1,4), (2,4)\}$, forming a pentagram when taken as edges of a graph; these edges then mark the contexts of the scenario. The vectors $|v_i\rangle$ considered in this work take the form
\begin{align}
|v_i \rangle &= \sin{(\alpha_i)}\cos{(\theta_i)}|11\rangle +\cos{(\alpha_i)}|10\rangle \nonumber\\[1ex]
&\qquad\qquad\qquad\quad+\sin{(\alpha_i)}\sin{(\theta_i)}|1-1\rangle
\end{align}
where $\{|j\, m\rangle:m=-j,\ldots, j\}$ is the Dicke basis of the $\hat{J}_z$ operator with the usual ordering. Three different sets of angles $\{(\alpha_i, \theta_i)\}_{i=0}^4$ are used to illustrate the influence that the choice of measurement operators has on the harvesting behaviour;  they are each given in Appendix  \ref{Appendix:QutritContextScenario} along with further details of their derivations. Since there are 5 contexts and each $\hat{B}_i$ only has outcomes $\pm 1$, there are $2^5=32$ possible global assignments and $5\times 2^2 = 20$ possible local assignments for this scenario, yielding a $20\times 32$ incidence matrix for the contextual fraction.

\section{Unruh-DeWitt Model} \label{Section:UdW}

To model quantum systems moving through spacetime that can interact with tensor quantum fields, the remarkably successful UDW model is turned to. One can think of the UDW model essentially as a quantum system following a trajectory in spacetime that can interact locally with a tensor quantum field around its trajectory. Despite its simplicity, the UDW model has been shown to capture important features of the interaction between quantum systems and quantum fields, such as the light-matter interaction \cite{lopp2021quantum} and the interaction with linearized quantum gravity \cite{Perche_Ragula_Martín-Martínez_2023}, to name just a couple. The model works as follows. 

Consider $D$ qudits with Hilbert spaces $\{\mathcal H_d\}_d^D$ and combined Hilbert space $\mathcal H = \bigotimes_d^D \mathcal H_d$, each moving along timelike trajectories $\mf z_d^\mu(\tau_d) = (t(\tau_d), x(\tau_d))\ , d \in\{1,\ldots,D\}$ in a globally hyperbolic spacetime $\mathcal{M}$ that admits a global timelike coordinate $t$, with the proper time for each qudit $\tau_d$ representing the Fermi normal coordinate time defined around each of their trajectories. Along each trajectory, each qudit has a free evolution governed by a free Hamiltonian $\hat{H}_{0_d}$, which is typically proportional to the energy gaps $\Omega^i_d \in \mathbb R_+\ , i \in \{1,\ldots,n\}$ between the $n$ energy levels of the qudits. Additionally, each qudit interacts with a background tensor quantum field $\hat{\mathcal O}(\mf x)$ along its trajectory that is prescribed by the interaction Hamiltonian density, in the interaction picture, as
\begin{equation}
    \hat{h}_d(\mf x)  = \lambda_d \Lambda_d(\mf x) \hat{\mu}_d(\tau_d) \otimes \hat{\mathcal O}(\mf x),
\end{equation}
where the $\lambda_d$ are coupling constants, $\hat{\mu}_d(\tau_d) = e^{i\hat{H}_{0_d}\Delta \tau}\hat{\mu}_de^{-i\hat{H}_{0_d}\Delta \tau}$ are interaction picture operators acting on the Hilbert space of qudit $d$, and $\Lambda_d(\mf x)$ are spacetime smearing functions, which define the shape of the interaction in spacetime. The functions $\Lambda_d(\mf x)$ are each strongly supported around the qudits' trajectories, $\mf z_d(\tau_d)$ and their spatial extent corresponds to the qudit's shape in their own rest frame.

The initial state of the entire system is $\hat{\rho}(t_0) = \hat{\rho}_{D}(t_0) \otimes \hat{\rho}_\phi$, with $\hat{\rho}_D(t_0) = \bigotimes_d^D \hat{\rho}_d(t_0)$ denoting the initial state of the qudits and $\hat{\rho}_\phi$ denoting the initial state of the tensor quantum field. Time evolution of this state to a final time $t$ is given by 
\begin{align}
\hat{\rho}(t) = \hat{U}(t, t_0)\hat{\rho}(t_0)\hat{U}^{\dagger}(t, t_0)\,,
\end{align}
with the time-evolution operators defined by
\begin{align} \label{Equation:TimeEvolve}
\hat{U}(t, t_0) = \mathcal{T}e^{\int^{t}_{t_0} \mathrm{d}V \hat{h}(\mf x)}\,,
\end{align}
where $\hat{h}(\mf x) = \sum_{i=1}^D \hat{h}_d(\mf x)$ is the Hamiltonian density of the full system and $\mathcal T$ is the time-ordering operator. The final time-evolved state can be obtained perturbatively from the Dyson expansion:
\begin{align}
\hat{\rho}(t) = \sum_{ij}\hat{\rho}^{(i,j)} = \sum_{ij}\hat{U}^{(i)}(t, t_0)\hat{\rho}(t_0)(\hat{U}^{(j)})^{\dagger}(t, t_0)
\end{align}
where
\begin{align}
\hat{U}^{(n)}(t, t_0) = \frac{(-i)^n}{n!}\prod^n_k \int \mathrm{d}V_{k+1} \mathcal{T}(\hat{h}(\mf x_{k+1}))\,.
\end{align}
The final state for the qudits can be computed by tracing out the field degrees of freedom from the final state: $\hat{\rho}_D(t) = \text{Tr}_\phi(\hat{\rho}(t))$, which can represent an entangled state between the qudits under certain circumstances. As is the case in much of the RQI literature \cite{Perche_2024}, it is assumed that the field starts out in a quasi-free state\footnote{A quasi-free state is a zero mean Gaussian state. Typical states that satisfy this condition are thermal states and most choices of vacua \cite{Haag}.}. 

The final state of the qudit system, up to second-order in the coupling constant $\lambda$, is then
\begin{align}
\hat{\rho}_D(t) = \hat{\rho}_D(t_0) + \hat{\rho}_D^{(1,1)} + \hat{\rho}_D^{(0,2)}+\hat{\rho}_D^{(2,0)} + \mathcal O(\lambda^4)
\end{align}
where
\begin{align} \label{Equation:RhoPert}
\hat{\rho}^{(1, 1)}_D &= \sum_{dd'}\lambda_d\lambda_{d'}\int\mathrm{d}V_1 \int \mathrm{d}V_2 \hat{\mu}_d(\tau_1)\hat{\rho}_D(t_0) \hat{\mu}_{d'}(\tau_2)\nonumber\\[1ex] 
&\times \Lambda_d(\mf x_1)\Lambda_{d'}(\mf x_2)W(\mf x_2, \mf x_1)\,, \\
\hat{\rho}^{(2, 0)}_D &= -\sum_{dd'}\frac{1}{2}\lambda_d\lambda_{d'}\int\mathrm{d}V_1 \int \mathrm{d}V_2 \Lambda_d(\mf x_1)\Lambda_{d'}(\mf x_2)\nonumber\\[1ex]
&\times\mathcal{T}\left(\hat{\mu}_d(\tau_1) \hat{\mu}_{d'}(\tau_2)W(\mf x_1, \mf x_2)\right)\hat{\rho}_D(t_0) \nonumber\\[1ex]
&= -\sum_{dd'}\lambda_d\lambda_{d'}\int\mathrm{d}V_1 \int \mathrm{d}V_2 \Lambda_d(\mf x_1)\Lambda_{d'}(\mf x_2)\nonumber\\[1ex]
&\times\hat{\mu}_d(\tau_1) \hat{\mu}_{d'}(\tau_2)\theta(\tau_1-\tau_2)W(\mf x_1, \mf x_2)\hat{\rho}_D(t_0)\,, \nonumber\\[1ex]
\hat{\rho}^{(0, 2)}_D &= (\hat{\rho}^{(2, 0)}_D)^{\dagger}\,,\nonumber
\end{align}

with $\theta(\tau_1-\tau_2)$ being the Heaviside step function and $W(\mf x_1,\mf x_2) = \langle \hat{\mathcal O}(\mf x_1)\hat{\mathcal O}(\mf x_2)\rangle_{\hat{\rho}_\phi}$ being the Wightman function, i.e. the correlation function of the field. If the field theory that is used obeys the commutation relation 
\begin{align}
\left[\hat{\mathcal O}(\mf x_1), \hat{\mathcal O}(\mf x_2)\right] = iE(\mf x_1, \mf x_2)
\end{align}
where $E(\mf x_1, \mf x_2) = G_R(\mf x_1, \mf x_2)-G_A(\mf x_1, \mf x_2)$ is called the causal propagator, with $G_R(\mf x_1, \mf x_2)$ and $G_A(\mf x_1, \mf x_2)$ being the retarded and advanced propagators of the field, respectively, then the Wightman function can be expressed in terms of a sum of the anti-commutator and commutator of the field
\begin{align}
W(\mf x_1, \mf x_2) &= \frac{1}{2}\left\langle \left \{\hat{\mathcal O}(\mf x_1), \hat{\mathcal O}(\mf x_2)\right\}\right\rangle_{\hat{\rho}_\phi}\nonumber\\[1ex]
&\qquad+\frac{1}{2}\left\langle \left[\hat{\mathcal O}(\mf x_1), \hat{\mathcal O}(\mf x_2)\right]\right\rangle_{\hat{\rho}_\phi}\nonumber \\[1ex]
&= \frac{1}{2}H(\mf x_1, \mf x_2) +\frac{i}{2}E(\mf x_1, \mf x_2)
\end{align}
where $H$ is called the Hadamard function. Since the field commutator is proportional to the identity operator of the field, it is clear that the causal propagator is independent of the field's state $\hat{\rho}_\phi$ because any state will commute with it. However, the causal propagator still encodes the state-independent degrees of freedom of the field as the commutator relation is fundamentally quantum. The Hadamard function on the other hand, contains all of the state-dependence (including the quantum degrees of freedom) of the field since different field states will not necessarily commute with the anti-commutator of the field.

In this paper, a 3+1-dimensional Minkowski spacetime is chosen such that $\tau = t$. The qudits are governed by a free Hamiltonian $\hat{H}_0 = \sum_d\hat{H}_{0_d}=\sum_d \Omega_d(\hat{J}^{(d)}_z+j_d\openone_d)$, where $j_d$ is the total angular momentum number for qudit $d$. The energy eigenstates are naturally represented in the Dicke basis $\{|j \, m\rangle_d: m=-j,\ldots, j\}$ with energy levels $\{0, \Omega_d,\ldots, 2j_d\Omega_d\}$, due to the shift of the ground state to 0 by the term $j_d\openone_d$. Each qudit has a monopole operator $\hat{\mu}_d = 2\hat{J}_x^{(d)}$, which yields, in the interaction picture, a time-dependent monopole operator given by
\begin{align}
\hat{\mu}_d(t) = e^{i\Omega_d t}\hat{J}^{(d)}_+ + e^{-i\Omega_d t}\hat{J}^{(d)}_-
\end{align}
where $\hat{J}^{(d)}_{\pm}$ are the usual $SU(2)$ ladder operators for each of the qudits. The qudits each couple to a massless scalar quantum field $\hat{\mathcal O}(\mf x) = \hat{\phi}(\mf x)$, which is a quantization of the scalar field $\phi(\mf x)$ whose dynamics are prescribed by the Klein-Gordon equation
\begin{align}
\nabla^\mu\nabla_\mu \phi(\mf x) = 0\,,
\end{align}
defining the associated Green's functions $G_R(\mf x_1, \mf x_2)$ and $G_A(\mf x_1, \mf x_2)$.

The quantum field $\hat{\phi}(\mf x)$ can be decomposed as
\begin{align}
\hat{\phi}(\mf x) = \int \frac{\mathrm{d}^3\bm k}{\sqrt{2(2\pi^3)|\bm k|}} \left(\hat{a}_{\bm k}^{\dagger} e^{i(|\bm k|t-\bm k \cdot \bm x)}+\hat{a}_{\bm k}e^{-i(|\bm k|t-\bm k \cdot \bm x)}\right)\,,
\end{align}
where $\hat{a}_{\bm k}, \hat{a}^{\dagger}_{\bm k}$ are the field ladder operators satisfying the commutation relation $[\hat{a}_{\bm k}, \hat{a}^{\dagger}_{\bm k'}] = \delta^3(\bm k - \bm k')$. The UDW system starts in the initial state $\hat{\rho}(t_0) = \hat{\rho}_D\otimes \hat{\rho}_\phi = \bigotimes_d^D|g\rangle_d \langle g| \otimes \hat{\rho}_\phi$, where $\hat{\rho}_\phi$ is chosen to be the vacuum state of the field $|0\rangle_\phi\langle 0|$ and $|g\rangle_d$ are the eigenstates of the spin-$(\text{dim}(\mathcal H_d)-1)/2$ total angular momentum operator in the $z$-direction $\hat{J}^{(d)}_z$ corresponding to the lowest value of $m$. An effective spacetime-smearing function 
\begin{align}
\Lambda^{\pm}_d(\mf x) &= \Lambda_d(\mf x)e^{\pm i\Omega_d t} \\
&= \sum_{l_d} \frac{ c_{l_d}e^{-(\bm x-\bar{\bm x}_{l_d})^T\bm \Sigma^{-1}_{l_d}(\bm x-\bar{\bm x}_{l_d})-\frac{(t-\bar{t}_d)^2}{T_d^2}\pm i\Omega_d t}}{\pi^{3/2} ||\tilde{c}_d||}\nonumber
\end{align}
will be used where $T_d \in \mathbb R_+$ is the temporal width for system $d$, $c_{l_d} \in \mathbb R$ are the basis coefficients of each basis function corresponding to system $d$, $\bar{t}_d \in \mathbb R$ is the temporal centre for system $d$, $\bar{\bm x}_{l_d} \in \mathbb R^3$ are the spatial centre vectors corresponding to each basis function for system $d$, and $||\tilde{c}_d|| = \sum_{m_d}c_{m_d}\sqrt{\text{det}\left(\bm \Sigma_{m_d}\right)}$. $\bm \Sigma_{l_d}$ is the 3-dimensional correlation matrix and is given by
\begin{align}
\bm \Sigma_{l_d} = \begin{pmatrix} \sigma_{1,l_d}^2 & \rho_{12, l_d} \sigma_{1,l_d}\sigma_{2,l_d} & \rho_{13, l_d}\sigma_{1,l_d}\sigma_{3,l_d} \\ \rho_{21, l_d}\sigma_{2,l_d}\sigma_{1,l_d} & \sigma_{2,l_d}^2 & \rho_{23, l_d}\sigma_{2,l_d}\sigma_{3,l_d} \\ \rho_{31, l_d}\sigma_{3,l_d}\sigma_{1,l_d} & \rho_{32, l_d}\sigma_{3,l_d}\sigma_{2,l_d} & \sigma_{3,l_d}^2\end{pmatrix}
\end{align}
where $\sigma_{i, l_d} \in \mathbb R_+$ is the spatial width corresponding to component $i$ ($i=1, 2, 3$) of each of the basis functions for system $d$ and $\rho_{ij, l_d} \in \mathbb R_+$ is the correlation between components $i$ and $j$. This smearing function allows for the representation of many spatial smearing profiles by simply choosing the appropriate values for the parameters of the Gaussian basis set (one must ensure that the normalization does not cause a blow-up of the function) \footnote{Note that one can also apply the same basis function expansion treatment to the temporal part of the basis, adding an extra sum over these indices, a second index on the expansion coefficients, and redefining the $||\tilde{c}_d||$ term to include the temporal widths. Additional generalizations could be to include polynomial terms $x^{l_{d,1}}y^{l_{d,2}}z^{l_{d,3}}$ to cover the full angular case. One also needs to define the centre of the resulting smearing profile since it can take an arbitrary shape.}. The form of the above smearing function will be useful for future studies, but for the purposes here, the sum will be taken to be over only one Gaussian basis function and  $\bm \Sigma_{l_d}$ is assumed to be diagonal with each element $ \sigma_{i, l_d}$ being the same. 

The smearing function $\Lambda^{p=\pm}_d(\mathsf x)$ will smear out the Wightman function and related bi-distributions (e.g. causal propagator, retarded and advanced propagators, Hadamard function, etc.), so it is useful to define the smeared version of an arbitrary bi-distribution $B(\mathsf x, \mathsf x')$ as
\begin{align}
B(\Lambda^{p}_{d}, \Lambda^{q}_{d'}) = \int\mathrm{d}V_1 \int \mathrm{d}V_2 \Lambda^{p}_d(\mf x_1)\Lambda^{q}_{d'}(\mf x_2)B(\mf x_1, \mf x_2)
\end{align}
and the forward time-ordered smeared version as 
\begin{align}
B_{\Delta t}(\Lambda_d^{p}, \Lambda_{d'}^{q}) = \int\mathrm{d}V_1& \int \mathrm{d}V_2 \Lambda^{p}_d(\mf x_1)\Lambda^{q}_{d'}(\mf x_2)\nonumber\\[1ex]
&\times \theta(t_1-t_2)B(\mf x_1, \mf x_2)\,,
\end{align}
where $\theta(t-t')$ is the Heaviside step function and the backward time-ordered smeared version $B_{-\Delta t}(\Lambda_d^{p}, \Lambda_{d'}^{q}) $ is just the above with the arguments of $B(\mf x_1, \mf x_2)$ and $\theta(t_1-t_2)$ reversed.

It is worth pointing out that non-point-like UDW models, such as the one considered here, explicitly break covariance by smearing the action of the monopole operator over multiple time slices such that the microcausality condition\footnote{$[\hat{\phi}(\mf x), \hat{\phi}(\mf x')] = 0$ for spacelike separated $\mf x, \mf x'$.} is violated by the interaction Hamiltonian, and also making the time-ordering reference frame dependent \cite{Martín-Martínez_Perche_Torres_2021,Perche_2024}. However, this only manifests in the perturbative terms beyond 2nd order provided that the initial state commutes with the free Hamiltonian or the smeared Wightman function goes to zero for spacelike-separated systems \cite{Perche_2024,Martín-Martínez_Perche_Torres_2021}, both of which are considered in our setup, so this kind of covariance violation will not appear in what follows. Smearing functions without compact support, like those considered here, also introduce further potential causality violations by allowing instantaneous signalling between spacelike separated regions, which is due to the fact that their support can extend across both regions \cite{deRamón_Papageorgiou_Martín-Martínez_2023,Martín-Martínez_2015}. One strategy to mitigate this problem is to define a temporal region that the smearing function is strongly supported on, which will be taken as
\begin{align}
-\eta_d T_d < t-\bar{t}_d < \eta_d T_d 
\end{align}
where $\eta_d$ is a positive constant that ensures strong temporal support.

Also note that because the free Hamiltonian is only a function of $\hat{J}_z$ in our setup, a non-contextual initial state could not be transformed into a contextual one under time-evolution with only the free Hamiltonian since $[\hat{\rho}(t_0), e^{-i\hat{H}_0\Delta t}] = 0$. For generic $e^{-i\hat{H}_0\Delta t}$ and $\hat{\rho}(t_0)$ one can see that the LHS of Eq. \eqref{Equation:NonContextInEq} is invariant with respect to transformations by $e^{-i\hat{H}_0\Delta t}$, so transformation to a contextual state from a non-contextual one using $e^{-i\hat{H}_0\Delta t}$ isn't possible in general.

The expressions for Eqs. \eqref{Equation:RhoPert} corresponding to the setup considered here can be found in Appendix \ref{Appendix:Calculations}.

\section{Contextuality Harvesting} 
\label{Section:ContextHarvest}

The core idea behind contextuality harvesting is conceptually similar to that of entanglement harvesting, although the former clearly demarcates the boundary between classical and non-classical quantum states, as mentioned in the introduction. In the standard entanglement harvesting scenario, two (or more) spatially separated UDW qudits that are both initially unentangled interact with a tensor quantum field for some time duration, leading to a joint final state for the qudits that corresponds to an entangled state after tracing out the field degrees of freedom. For the basic contextuality harvesting scenario, there is only one UDW qudit that is initially in a non-contextual state, which then interacts with a tensor quantum field for some time duration, leading to a final state that is contextual according to a genuine contextuality measure such as the contextual fraction. 

The entanglement harvesting protocol includes the term 'harvesting' because entanglement between systems interacting via a mediator system cannot be generated through only local operations and classical communication~\cite{horodecki2009quantum} -- there had to have been pre-existing entanglement in the mediator system for this to be possible. One can also argue that contextuality is really ``harvested" from the state of the quantum field by similarly establishing a resource theory for contextuality and using the fact that it is already known that quantum field states harbour non-classicality (predominantly in the form of non-locality), which References ~\cite{Summers_Werner,Summers_Werner_1985,Summers_Werner_1995, Summers_Werner_1987a,Summers_Werner_1987b,Summers_Werner_1987c,Kitajima_2006,Kitajima_2017,Schlichtholz_Markiewicz_2024,Schlichtholz_Mandarino_Żukowski_2022,Benatti_Floreanini_Narnhofer_2025,Butterﬁeld_Halvorson,Clifton_Halvorson_2001,DeFabritiis_Roditi_Sorella_2023,DeFabritiis_Guedes_Guimaraes_Roditi_Sorella_2024,Grossi_Barata_2024,Guedes_Guimaraes_Roditi_Sorella_2024,Valente_2014,Rédei_1989,Rédei_1991} show is quite a generic feature that most infinite-dimensional states possess -- including the vacuum state. Now in the resource theory of contextuality, the set of free operations that preserve the non-contextual state space for a given contextuality scenario are defined to essentially be relabelings of the measurements/outcomes and coarse-grainings of measurement sets/outcomes ~\cite{Amaral_2019}. These operations are free because they cannot introduce more measurement incompatibility into the measurement set, which is basically akin to increasing contextuality. This means that for a fixed scenario, a non-contextual initial state cannot become contextual under the free operations alone. If one decides to change the scenario by introducing more incompatibility, they can of course convert the non-contextual initial state to a contextual one, but then this operation is not free and the emerging non-classicality is explained by this non-free operation. One can also establish a contextuality resource theory for the field and field plus probe -- using a continuous-variable analogue of the contextual fraction for the contextuality monotone~\cite{Barbosa_Douce_Emeriau_Kashefi_Mansfield_2022} -- which will look similar to that of the probe system but with many more states counting as useful resources. Drawing on the link between magic and contextuality discussed near the end of the introduction, many families of operations can be classified as free operations, including Clifford circuits on odd-prime-dimensional Hilbert spaces  and Gaussian operations in the continuous-variable setting. Due to the ubiquity of non-classicality in quantum field states, transformations acting on the joint probe-field system will tend to preserve this non-classicality and transfer some of it to the probe system through correlations such as entanglement. 

Operations which do not involve interacting with the field may still increase the contextuality of just the probe system, but the contextuality resource theory for the probe can be used to identify such operations so that the increase of contextuality witnessed can be explained. It is the three resource theories for the probe, field, and probe-field together, that allow one to certify the source of any contextuality acquired. This is analogous to how the argument for entanglement harvesting based on the resource theory of Local Operations and Classical Communication goes through, but with a different set of free operations and resources. 

Therefore, if the contextuality scenario is fixed, the probe system's state is initialized to a non-contextual state with respect to the scenario, and an odd prime dimensional qudit state (a qutrit in this work) is used, then given the pervasiveness of non-classicality in quantum fields, it is reasonable to claim that the contextuality generated by this interaction was harvested from the quantum field through this protocol, which is the stance taken in this work. A more rigorous argument will be the subject of future work.


The contextuality measure that shall be used to quantify the amount of contextuality and/or non-local entanglement harvested will be the contextual fraction difference between an initial system state $\hat{\rho}_D(t_0)$ and a final system state $\hat{\rho}_D(t)$:
\begin{align}
\Delta CF(\bm v^e) = CF(\bm v_{t}^e)-CF(\bm v_{t_0}^e)\,.
\end{align}
where $\bm v^e_t$ is the vectorized version of the empirical model corresponding to the final state, which has components 
\begin{align}
(\bm v^e_t)_i = \text{Tr}\left(\hat{\rho}_D(t)\hat{P}_i(t)\right)\,,
\end{align}
where the index $i$ ranges over all of the (tensor) products of eigenprojectors (in a context) corresponding to the interaction picture measurement operators in the scenario. 

The contextual fraction of a given empirical model corresponding to a specific scenario quantifies the fraction of the model that can be explained by a strongly contextual model \cite{Abramsky_Barbosa_Mansfield_2017}. So given a particular state, this tells you how much the description of the state in one context conflicts with its description in other contexts. In the case of non-local entanglement, it is the degree to which the non-locality arising from subsystem interactions alters the local descriptions of those subsystems. For the setups that are considered, taking the difference between the contextual fractions before and after the interaction can be understood as subtracting away the initial contextuality of the empirical model (i.e. contextuality in the absence of any transformation) and quantifying the influence of only the quantum field on disrupting the consistency of the state's description in different contexts. 


Since the goal is to harvest contextuality from the quantum degrees of freedom of the vacuum state of the quantum field, rather than acquiring contextuality through field-mediated signalling, a way to distinguish the two is necessary. As stated in reference \cite{Perche_2024}, the Feynman propagator can be formed from the Wightman function as
\begin{align} \label{Equation:FeynProp}
G_F(\mf x, \mf x') = \theta(t-t')W(\mf x, \mf x')+\theta(t'-t)W(\mf x', \mf x)\, ,
\end{align}
where $\theta(t-t')$ is the Heaviside step function, and note that $G_F$ can be broken down into a sum of two functions 
\begin{align}
G_F(\mf x, \mf x') = \frac{1}{2}H(\mf x, \mf x') +i\frac{1}{2}\Delta(\mf x, \mf x')
\end{align}
where $H$ is again the Hadamard function and $\Delta$ is a measure of causal contact called the symmetric propagator. $\Delta$ can be expressed in terms of $G_R$ and $G_A$ as $\Delta(\mf x, \mf x') = G_R(\mf x, \mf x')+G_A(\mf x, \mf x')$, or alternatively as $\Delta(\mf x, \mf x') = 2G_R(\mf x, \mf x')-E(\mf x, \mf x')$ from the expression $G_A(\mf x, \mf x') = G_R(\mf x, \mf x')-E(\mf x, \mf x')$. The author of reference ~\cite{Perche_2024}, then outlines the conditions for genuine harvesting of entanglement as
\begin{align}\label{Equation:NegHarvestConds}
\frac{1}{2}\left|\Delta(\Lambda_d^{+}, \Lambda_{d'}^{+})\right| \ll \mathcal N(\hat{\rho})\ \ \text{and} \ \ \mathcal N(\hat{\rho}) > 0 \,,
\end{align}
with the negativity entanglement measure $\mathcal N(\hat{\rho})$ given by
\begin{align}
\mathcal N(\hat{\rho}) &= \left|\sum_{\lambda_i < 0} \lambda_i\right|\,,
\end{align}
where $\lambda_i$ are the eigenvalues of the partially-transposed density operator $\hat{\rho}^{T_d}$. The first condition in Eq. \eqref{Equation:NegHarvestConds} is motivated by the fact that the smeared symmetric propagator is a measure of causal contact between systems, meaning that it can essentially be taken as a measure of how little the quantum degrees of freedom of the quantum field are being used to facilitate the increase of  entanglement between the systems. The choice of using the $++$ component is a consequence of the far-off-diagonal terms that appear in the density matrix they consider, together with the definition of entanglement. How the smeared symmetric propagator scales in comparison to the smeared Hadamard function is also of interest to us and will help identify good parameter regimes as well.

Based on the expressions for $\hat{\rho}_D(t)$ in Appendix \ref{Appendix:Calculations}, the conditions that are of interest here are that the smeared Hadamard term dominates over the smeared symmetric propagator term, and that the contextual fraction difference is greater than 0, i.e.
\begin{align} \label{Equation:ContextHarvestConds}
\left|\frac{\Delta(\Lambda_d^{+}, \Lambda_{d'}^{+})}{H(\Lambda_d^{+}, \Lambda_{d'}^{+})}\right| \ll 1\ \ \text{and} \ \ \Delta CF(\bm v^e) > 0 \,.
\end{align}
The first condition is necessary due to the possibility of increasing the contextuality of a system through field-mediated signalling of measurement results, which was shown in \cite{Porto_Ruffolo_Rabelo_Cunha_Kurzynski_2024} for a similar non-relativistic qubit-qutrit scenario. If these conditions are met then one can be reasonably sure that the harvested contextuality is predominantly due to the preexisting quantum correlations in the field and not because of field-mediated signalling. 



The above conditions for genuine contextuality harvesting are quite general, applying to many different spacetimes and quantum fields. Particularizing to our setting of a massless scalar quantum field in 3+1 Minkowski spacetime, the first condition combined with the expressions derived in the appendix for $\Delta(\Lambda_d^{+}, \Lambda_{d'}^{+})$ and $H(\Lambda_d^{+}, \Lambda_{d'}^{+})$ (equations \eqref{Equation:SymProp}
and \eqref{Equation:Hadamard}, respectively) already reveal much information about how conducive this setting is to harvesting. For instance, observe the case when $d=d'$ and each of the sums in the two expressions has only one element:
\begin{align}\label{Equation:FirstHarvestCondSame}
&\left|\frac{\Delta(\Lambda_d^{+}, \Lambda_{d}^{+})}{H(\Lambda_d^{+}, \Lambda_{d}^{+})}\right| = \sqrt{\frac{2T_d^2}{\beta_{1_d1_d}}} = \frac{T_d}{\sigma_{1_d}}\,.
\end{align}
The ratio of the temporal to spatial smearing width is the only determining factor for harvesting under the first condition, independent of the internal details of the UDW systems. In particular, notice that harvesting is only possible if the temporal smearing width is small while the spatial smearing width remains fixed at some intermediate value such that both are within the region of strong support, the spatial smearing width is large while the temporal smearing width remains fixed at some intermediate value (again, in the strongly supported region), or if one is a scalar multiple of the other such that the coefficient is less than one. In this last case, harvesting is permitted when the spacetime smearing becomes highly localized, but not when it becomes highly dispersed since this would exceed the region where the smearing function is strongly supported (introducing non-physical signalling). 

The case when $d \neq d'$ has a richer structure that requires further analysis to elucidate fully. Let us assume again that each of the sums has only one element and furthermore that the spatial smearing widths and system energy gaps for each of the systems are equal. The first condition then reduces to 
\begin{widetext}
\begin{align}\label{Equation:SymHadRatDiff}
&\left|\frac{\Delta(\Lambda_d^{+}, \Lambda_{d'}^{+})}{H(\Lambda_d^{+}, \Lambda_{d'}^{+})}\right| = \left|\frac{e^{L_{l_dm_{d'}}a_1}\text{erfi}(a_3-ia_2L_{l_dm_{d'}})+e^{-L_{l_dm_{d'}}a_1}\text{erfi}(a_3+ia_2L_{l_dm_{d'}})}{e^{L_{l_dm_{d'}}a_1}\text{erfi}(b_2+b_1L_{l_dm_{d'}})-e^{-L_{l_dm_{d'}}a_1}\text{erfi}(b_2-b_1L_{l_dm_{d'}})}\right|
\end{align}
\end{widetext}
where 
\begin{align*}
b_1 &= \frac{1}{\sqrt{\beta_{1_d1_d}+T_d^2+T_{d'}^2}}\,, \nonumber\\[1ex]
a_1 &= b_1^2(i\Omega_d(T_d^2-T_{d'}^2)+2\Delta \bar{t}) \,,\nonumber\\[1ex]
b_2 &= -\frac{a_1}{2b_1}\,, \nonumber\\[1ex]
a_2 &= \frac{b_1\sqrt{T_d^2+T_{d'}^2}}{\sqrt{\beta_{1_d1_d}}}\,, \nonumber\\[1ex]
a_3 &= -\frac{ia_1}{a_2\beta_{1_d1_d}}\,, \nonumber\\[1ex]
\end{align*}
and $L_{l_dm_{d'}}, \Delta \bar{t}$ are defined in Appendix \ref{Appendix:Calculations} as the distance between the centres of the spatial/temporal smearing regions corresponding to systems $d$ and $d'$, respectively. One can now see that the above ratio tends to 0 for small $\sigma_{1_d}$, or small $T_d,T_{d'}$ and $\Delta \bar{t} = 0$, assuming the other parameters in the denominator are chosen such that it stays non-zero, indicating that sharply peaked spatial or temporal smearing distributions constitute regimes that only rely on existing field correlations for the harvesting. Contrast this with Eq. \eqref{Equation:FirstHarvestCondSame}, which shows that harvesting for small $\sigma_{1_d}$ is possible for both cases if $T_d$ and/or $T_{d'}$ are also small. If the temporal smearing widths are not small then harvesting will only happen for the $d\neq d'$ case. Similarly, if the temporal smearing widths are small (and $\Delta \bar{t}=0$) and $\sigma_{1_d}$ does not approach 0, then harvesting will be possible in both cases; note also that as $\sigma_{1_d}$ gets larger, harvesting in the $d=d'$ case becomes more likely while it decreases in the $d\neq d'$ case. Intermediate parameter regimes likely exist where more of the interplay between local and non-local contextuality harvesting occurs, which is left until sections \ref{SubSection:SingleQutrit} and \ref{SubSection:QubitQutrit}. However, one intermediate regime where genuine harvesting is not possible in either case happens when $T_d=T_{d'}=\sigma_{1_d}=1$ and $\Delta \bar{t}=L_{l_dm_{d'}}$: here the harvesting condition for $d=d'$ is clearly violated and for $d\neq d'$, the condition is
\begin{align}
&\left|\frac{\Delta(\Lambda_d^{+}, \Lambda_{d'}^{+})}{H(\Lambda_d^{+}, \Lambda_{d'}^{+})}\right| = e^{L_{l_dm_{d'}}^2}\left|\frac{\text{erf}(L_{l_dm_{d'}})}{\text{erfi}(L_{l_dm_{d'}})}\right|
\end{align}
which is lower-bounded by 1 and diverges for larger $L_{l_dm_{d'}}$, thus also violating the condition. 

Looking at Eq. \eqref{Equation:SymHadRatDiff} again, both the numerator and denominator go to 0 for large $\sigma_{1_d}$ or large $T_d$ and $ T_{d'}$, assuming the other parameters are chosen accordingly. These limits for the spatial and temporal smearing correspond to spreading out the interaction regions well beyond the strongly supported zones to eventually cover the whole region, which is a highly non-local interaction that is clearly unphysical. One can also take the very small $L_{l_dm_{d'}}$ limit to see that the numerator converges to $2\,\text{erfi}(a_3)$ while the denominator converges to 0, so that the ratio diverges and no harvesting is possible. In the opposite limit, the erfi terms in the numerator are evaluated at $\pm i\infty$  which converts them to standard error functions evaluated at $\pm \infty$, yielding $-1$ for the first and $+1$ for the second. The decaying exponential in the second term sets that term to 0, eventually leaving
\begin{align}
& \lim_{L_{l_dm_{d'}}\to \infty}\Big|\text{erfi}(b_2+b_1L_{l_dm_{d'}})\nonumber\\[1ex]
&\qquad\qquad\qquad -e^{-2L_{l_dm_{d'}}a_1}\text{erfi}(b_2-b_1L_{l_dm_{d'}})\Big|^{-1}
\end{align}
which converges to 0 as long as the other parameters are chosen accordingly. Consequently, it would seem that harvesting is allowed in this regime. However, it is clear from Eq. \eqref{Equation:FeynProp} that the smeared Hadamard function and the smeared symmetric propagator both actually converge to 0, so one must interpret this result as signifying that the smeared symmetric propagator converges to zero more rapidly than the smeared Hadamard function does. 

In the following, two different setups will be considered: 1) a single qutrit UDW system harvests contextuality and 2) a qubit-qutrit UDW system harvests contextuality and entanglement. In the former case, the point is to consider a minimal example where the UDW system can only exhibit local contextuality and to illustrate the general contextual nature and capacity to harbour non-classical resources of the vacuum state of the quantum field. In the latter case, the interplay between local contextuality and entanglement along with their robustness as quantum resources will also be studied. The code to reproduce the calculations and plots is available in a public GitHub repository ~\cite{ContextHarvest_Github}.


\subsection{A Single UDW Qutrit} \label{SubSection:SingleQutrit}

A single system can only have entanglement if it admits a tensor product decomposition of its Hilbert space into smaller Hilbert spaces such that $\text{dim}(\mathcal H)=\prod^D_d \text{dim}(\mathcal H_d)$. This is clearly impossible for $p$-dimensional Hilbert spaces, where $p$ is a prime number, since by definition a prime number cannot be expressed only as a product of factors. Recalling that contextuality can only be observed in systems with Hilbert space dimension greater than 3, a single qutrit system is then the minimal system to witness only contextuality and no  entanglement. Note that recently it was shown that a single qutrit could acquire magic from interacting with a massless scalar quantum field \cite{Nyström_Pranzini_Keski-Vakkuri_2024}, which is a type of non-classical resource used in quantum computation that is powered by contextuality in many cases~\cite{Howard_Wallman_Veitch_Emerson_2014,Delfosse_Okay_Bermejo-Vega_Browne_Raussendorf_2017}\footnote{The equivalence holds for qutrit systems, which is what is used here and in Reference ~\cite{Nyström_Pranzini_Keski-Vakkuri_2024}.}. However, as noted earlier, contextuality is a much more general non-classical resource that can encompass magic and non-local entanglement as subsets, so it is still worthwhile to demonstrate that this larger family of resources can be harvested.

For a single qutrit, $\hat{J}_z$ is the spin-1 angular momentum operator in the $z$-direction, with eigenstates represented in the Dicke basis as $\{|j=1\, m\rangle\}_{m=-1}^1$ with the usual ordering, and a ground state $|g\rangle = |-1\rangle$. The basis elements will be abbreviated as $|m\rangle$ from now on. Using the expressions from Appendix \ref{Appendix:Calculations}, the final density matrix representation of our single qutrit system in our prescribed basis is given by
\begin{align}
\hat{\rho}_D(t) = \begin{pmatrix} 0 & 0 & -2\bar{W}^{++}_{11} \\ 0 & 2W^{+-}_{11} & 0\\ -2(\bar{W}^{*})^{++}_{11} & 0 & 1-2W^{+-}_{11}\end{pmatrix} + \mathcal O(\lambda^4)
\end{align}

where $\bar{W}^{pq}_{dd'} = \lambda_d\lambda_{d'}W_{\Delta t}(\Lambda_d^{p}, \Lambda_{d'}^{q})$ and $W^{pq}_{dd'} = \lambda_d\lambda_{d'}W(\Lambda_d^{p}, \Lambda_{d'}^{q})$. Here it can be seen that the field state-dependence is distributed throughout the density matrix, with the off-diagonal elements containing the relevant terms for the genuine harvesting condition. Indeed, the choice of the $++$ component in the harvesting conditions can be motivated by its function in the non-contextuality inequality LHS of the modified pentagram scenario, since the absence of these terms in the final state would yield a state that would not violate the inequalities and hence be considered non-contextual. Seeing as how the off-diagonal terms also represent the acquired coherence of the state, it is tempting to try and give the interpretation that the amount of harvested contextuality is about the tradeoff between the coherence terms and the ``local" noise terms; this is, however, highly dependent on the given contextuality scenario. For example, the $|1\rangle$ eigenstate of the spin-1 $\hat{J}_z$ operator -- having 0 coherence -- maximally violates the non-contextuality inequality in the typical KCBS scenario that was introduced earlier in Section \ref{Section:Contextuality}. Fortunately, the final state and contextuality scenario used in this work does admit the coherence interpretation, so one may use either to make sense of the results.


\begin{figure*}[t!]
    \includegraphics[width = 1 \linewidth]{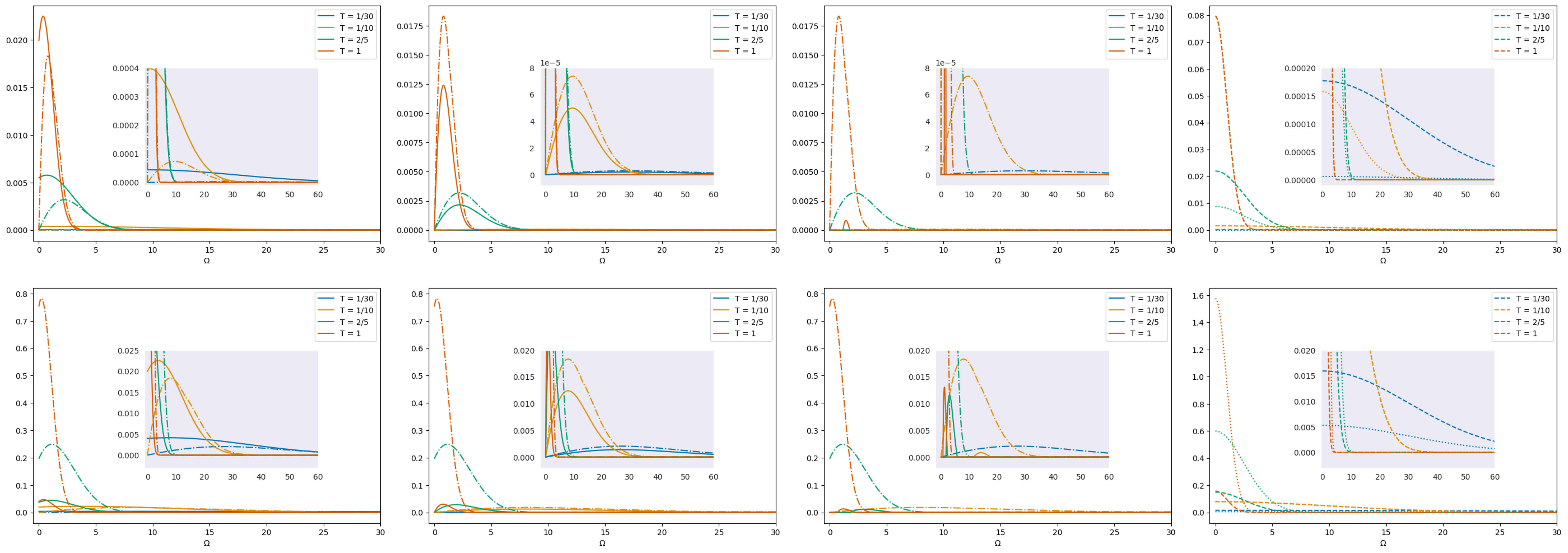}
    \caption{Plots for the single qutrit harvesting scenarios. The first three columns of plots depict $\Delta CF(\bm v^e)/\lambda^2$ with solid curves and $M(\hat{\rho})/\lambda^2$ with dot-dashed curves, while the last depicts $|\Delta(\Lambda_d^{+}, \Lambda_{d}^{+})|$ with dotted curves and $|H(\Lambda_d^{+}, \Lambda_{d}^{+})|$ with dashed curves, all as functions of the energy gap $\Omega$ for different temporal smearing widths and a coupling constant of $\lambda = 10^{-4}$. The top plots correspond to a spatial smearing width of $\sigma_{1_1} = 1$, while the bottom plots correspond to $\sigma_{1_1} = 0.1$. All plots additionally consider $\bar{t}_d=0$. The columns of the first three columns of plots represent the 3 sets of angles for the scenarios introduced in Appendix \ref{Appendix:QutritContextScenario}, with the order of the columns following the order of appearance of the sets of angles.}
    \label{Figure:CFDManaSingleQutrit}
\end{figure*}

Figure \ref{Figure:CFDManaSingleQutrit} depicts $\Delta CF(\bm v^e)/\lambda^2$ and $M(\hat{\rho})/\lambda^2$ in the first three columns of plots corresponding to each of the three sets of measurement operators in the modified pentagram scenario in Appendix \ref{Appendix:QutritContextScenario}, while the last column of plots depicts $|\Delta(\Lambda_d^{+}, \Lambda_{d}^{+})|$ and $|H(\Lambda_d^{+}, \Lambda_{d}^{+})|$, all as functions of the energy gap $\Omega_d$ for various values of $\sigma_{l_d}$ and $T_d$. Observe in the first three columns of plots that for larger spatial smearing widths (top plots), the harvested contextuality is smaller in magnitude, generally widening and decreasing in magnitude as the temporal smearing width decreases. For decreasing spatial smearing widths, the trend continues. These observations make sense, as configurations with larger spatial smearing weaken the harvesting magnitude by spreading out the interaction more, and temporal smearing profiles that are sharper than the spatial smearing will cause the qutrit to spend less time at peak interaction strength for spatial regions with non-negligible interaction strength, thus the qutrit could not harvest as much contextuality as it could with a less sharply peaked temporal smearing profile relative to the spatial smearing profile.

The last column of plots brings to our attention the fact that not all of the settings in Figure \ref{Figure:CFDManaSingleQutrit} correspond to genuine contextuality harvesting according to the conditions set out by Eq. \eqref{Equation:ContextHarvestConds}. Indeed, it is only really the first three $T_d$ cases in the first row of plots and the $T_d=1/30$ case in the second row of plots that satisfy them, which one can see from the much larger magnitude of the smeared Hadamard function (dashed curve) compared with the smeared symmetric propagator (dotted curve) over the peaked regions of the corresponding contextual fraction difference. As the temporal smearing width increases one sees that the smeared symmetric propagator begins to completely dominate the smeared Hadamard function for the significant regions of the contextual fraction difference, and thus the system is relying less on the quantum degrees of freedom of the field to acquire contextuality. This is consistent with the fact that making the interaction longer will better allow for field-mediated signalling. 

Another surprising observation from the plots in the first column of Figure \ref{Figure:CFDManaSingleQutrit} is that the contextual fraction difference corresponding to $\Omega_d=0$ in all cases is nonzero. This can be confirmed by looking at the expression for the difference between the LHS of the non-contextuality inequality corresponding to the modified pentagram scenario for the final and initial state:
\begin{align}\label{Equation:QutritContextIneq}
S_{\mathcal C} &= \frac{T_1^2\lambda_1^2e^{-\frac{T_1^2 \Omega_1^2}{2}}}{\pi \left(2 T_1^2 + 2\sigma_1^2\right)^{\frac{3}{2}}}\Bigg(\ell_1\sqrt{\pi} T_1^2 \Omega_1 \Bigg(\nonumber \\[1ex]
&1-\text{erf}\left(\frac{T_1^2 \Omega_1}{\sqrt{2 T_1^2 + 2\sigma_1^2}}\right)\Bigg)e^{\frac{T_1^4 \Omega_1^2}{2T_1^2 + 2\sigma_1^2}}\nonumber\\[1ex]
&-(\ell_1-2\ell_2) \sqrt{2T_1^2 + 2\sigma_1^2}\Bigg)\,,
\end{align}
where the coefficients $\ell_1$ and $\ell_2$ come from the measurement operators associated with the scenario. Here one can directly see that only the last term survives when setting $\Omega_d = 0$ provided $\ell_1 \neq 2\ell_2$.  In fact, as long as $\ell_2$ is sufficiently large, any admissible value for $\ell_1$ will allow non-zero harvesting for $\Omega_d = 0$, as can be seen by plotting Eq. \eqref{Equation:QutritContextIneq} for various combinations of parameters. This is in stark contrast with many results in the literature that demonstrate how gapless UDW systems cannot harvest entanglement \cite{Perche_2024}, which distinguishes contextuality as a resource and highlights the fact that contextuality offers an alternative route for the presence of nonclassicality in the system. The presence of this feature, however, depends on the given contextuality scenario and the specific choice of measurement operators; this is made clear in the second column of plots, where the harvested contextuality is 0 for $\Omega_d = 0$. For the three sets of angles used in this work, the constants $\ell_1, \ell_2$ have the following values displayed in Table \ref{Table:ContextInEqConsts}.
\begin{table}
    \centering
    \begin{tabular}{|c|c|l|}\hline
 $S_{\mathcal C}$ for $\Omega_d = 0$& $\ell_1$& $\ell_2$\\\hline
 $> 0$& ~$ 0.390345$& ~$ 0.445589$\\\hline
 $= 0$& ~$ 2.174898$& ~$ 0.537879$\\\hline
 $< 0$& ~$ 0.902216$& ~$ 0.451108$\\ \hline\end{tabular}
    \caption{Table of values for the parameters $\ell_1, \ell_2$ appearing in $S_{\mathcal C}$.}
    \label{Table:ContextInEqConsts}
\end{table}

The first three columns of plots of Figure \ref{Figure:CFDManaSingleQutrit} also offer a comparison between harvested magic and contextuality by plotting the contextual fraction difference with the mana expression Eq. \eqref{Equation:ManaUDW} from Appendix \ref{Appendix:ManaExpression} for the three sets of angles from Appendix \ref{Appendix:QutritContextScenario}. The contextual fraction difference is seen to have a similar profile to the mana but also seems to indicate that one can harvest more non-classical resources over a similar range of energy gaps in certain parameter regimes and with specific measurement sets than is suggested by the mana -- namely, the setup corresponding to the first column of plots. In particular, nonclassicality in the form of contextuality persists in the system even when the amount of magic is 0. For other measurement sets (second and third columns of plots), it appears that the amount of harvested contextuality is less than that of magic -- the magic displayed in the third column in particular, towers over the harvested contextuality. These observations are perhaps unsurprising since magic is a special type of contextuality whose state space boundary is with respect to stabilizer states, which need not coincide with that of an arbitrary contextuality scenario -- meaning there will be many more states in general that are contextual but not magical. The fact that the final system state in many parameter regimes possesses non-classicality with respect to the scenario considered in this work, the set of stabilizer operations, and negativity further highlights the work done by the non-classicality of the quantum field. Finally, the first three plots in the bottom row of Figure \ref{Figure:CFDManaSingleQutrit} ($\sigma_{1_1} = 0.1$ and $T_1$ large enough) demonstrate that magic can also be harvested for gapless systems under certain parameter regimes. 

\subsection{A Qubit-Qutrit UDW System} \label{SubSection:QubitQutrit}

According to recent work in the non-relativistic regime \cite{Xue_Xiao_Ruffolo_Mazzari_Temistocles_Cunha_Rabelo_2023,Porto_Ruffolo_Rabelo_Cunha_Kurzynski_2024}, there exist scenarios where adding an additional qubit system to our qutrit system from before introduces the possibility of harvesting entanglement, contextuality, both, or neither. To investigate this further, the negativity will be computed for a qubit-qutrit system and compared with the harvested contextuality of the qutrit using the scenarios from Subsection \ref{SubSection:SingleQutrit}. The idea is that the modified pentagram scenario constitutes a local contextuality witness for only the qutrit system, whereas the negativity serves as an entanglement witness for the full system. Thus, since the two measures use the components of the state differently, a tradeoff relationship can emerge where certain parameter regimes favour local contextuality over entanglement and vice-versa.

With the above setup in mind, there are several interesting features of the entanglement-contextuality relationship that will be further elucidated here. 
\begin{enumerate}
    \item Which parameter regimes allow for harvesting of entanglement, contextuality, both, or neither?
    \item How does contextuality affect the amount of harvested entanglement and vice versa?
\end{enumerate}

The first question already has a partial answer from the analysis at the end of Section \ref{Section:ContextHarvest}, but now the second harvesting condition must be tested to provide complete answers. That analysis plus the expressions from the appendix also supply us with some intuition for the second question as to how the two quantities influence each other, but again, the negativity needs to be calculated to have a complete picture. This is what will now be done in the following.

For a qubit-qutrit system, $\hat{J}_z^{(d)}$ represents the spin-1/2 and spin-1 angular momentum operators in the $z$-direction, respectively. $\hat{J}_z^{(1)}$ has eigenstates $\{|-1/2\rangle, |1/2\rangle\}$ and a ground state $|g\rangle_1 = |-1/2\rangle$, while $\hat{J}_z^{(2)}$ has eigenstates $\{|m\rangle\}_{m=-1}^1$ and a ground state $|g\rangle_2 = |-1\rangle$. Using the expressions from Appendix \ref{Appendix:Calculations}, the final density matrix representation of our qubit-qutrit system in the basis $\{|1/2\rangle \otimes |1\rangle, |1/2\rangle \otimes |0\rangle, |1/2\rangle \otimes |-1\rangle, |-1/2\rangle \otimes |1\rangle, |-1/2\rangle \otimes |0\rangle, |-1/2\rangle \otimes |-1\rangle\}$ is given by
\begin{align} \label{Equation:QubitQutritDen}
\hat{\rho}_D(t) =  \begin{pmatrix} 0 & 0 & 0 & 0 & 0 & 0 \\ 0 & 0 & 0 & 0 & 0 & r^{*}_{62} \\ 0 & 0 & W^{+-}_{11} & 0 & \sqrt{2}W^{+-}_{21} & 0 \\ 0 & 0 & 0 & 0 & 0 & r_{46} \\ 0 & 0 & \sqrt{2}W^{+-}_{12} & 0 & 2W^{+-}_{22} & 0 \\ 0 & r_{62} & 0 & r^{*}_{46} & 0 & r_{66}\end{pmatrix} + \mathcal O(\lambda^4)
\end{align}

$r_{66} = 1-W^{+-}_{11} - 2W^{+-}_{22}$, $r_{62} = -\sqrt{2}(G^{*})^{++}_{21}$, $
r_{46} = -2\bar{W}^{++}_{22}$, and $G^{pq}_{dd'} = \lambda_d\lambda_{d'}G_F(\Lambda^p_d, \Lambda^q_{d'})$. One can again see the most relevant contributions from the field state-dependence in the smeared Hadamard portion of the off-diagonal terms spread across the combined system, with the $r_{46}$ terms directly influencing the harvested qutrit contextuality and the $r_{62}$ terms having more of an impact on the harvested entanglement of the joint system. These terms also carry with them the only contributions from classical signalling through the smeared symmetric propagator, and so will be the most important with respect to the first condition of Eq. \eqref{Equation:ContextHarvestConds} in determining what can be genuinely harvested. The above density matrix makes it explicitly clear how the entanglement of the full system and the contextuality of the qutrit system can jointly influence the amount of contextuality/entanglement that can be harvested: the parameters appearing in the non-local terms are the same as those in the local terms and each set of terms has a different profile, meaning that different and potentially conflicting parameter regimes will optimize each term.

The reduced final state for just the qutrit system is then given by taking the partial trace of $\hat{\rho}_D(t)$ with respect to the qubit system:
\begin{align} \label{Equation:RedDen}
\hat{\rho}_{2}(t) = \begin{pmatrix} 0 & 0 & -2\bar{W}^{++}_{22} \\ 0 & 2W^{+-}_{22} & 0 \\ -2(\bar{W}^{*})^{++}_{22} & 0 & 1-2W^{+-}_{22}\end{pmatrix}  + \mathcal O(\lambda^4)
\end{align}

which has the same form as the single qutrit density matrix from Subsection \ref{SubSection:SingleQutrit}.

The density matrices Eqs. \eqref{Equation:RedDen} and \eqref{Equation:QubitQutritDen} offer a concrete visualization of the relationship between entanglement and contextuality in the relativistic regime by illustrating how the entanglement and contextuality are distributed throughout the system and in which terms specifically. Because each of the terms has a physical meaning, this also provides a nice mechanistic explanation for how the non-classicality arises in the first place. For instance, the smeared Feynman propagator terms $r_{62}$ contain a contribution from field-mediated classical communication between subsystems, but also contain contributions from the vacuum correlations of the field along with the smeared Wightman function terms that appear throughout the state. 

The eigenvalues of the partially-transposed joint density matrix Eq. \eqref{Equation:QubitQutritDen}, with respect to either the qubit or qutrit system, allow for the negativity to be computed up to second-order in the coupling constant $\lambda$:
\vspace{2em}
\begin{widetext}
\begin{align}
\mathcal N(\hat{\rho}_D(t)) &= \frac{1}{2}\bigg|\min \Big(0, \sqrt{(1 - W^{+-}_{11} - 2W^{+-}_{22})^2 + 8W^{+-}_{21}W^{+-}_{12} + 4|r_{46}|^2} + 1 - W^{+-}_{11} - 2W^{+-}_{22}\Big)\nonumber\\
&\quad +\min \Big(0, -\sqrt{(1 - W^{+-}_{11} - 2W^{+-}_{22})^2 + 8W^{+-}_{21}W^{+-}_{12} + 4|r_{46}|^2} + 1 - W^{+-}_{11} - 2W^{+-}_{22}\Big)\nonumber \\
& \quad +\min \Big(0, \sqrt{(W^{+-}_{11} - 2W^{+-}_{22})^2 + 4|r_{62}|^2} + W^{+-}_{11} + 2W^{+-}_{22}\Big) \nonumber \\
& \quad +\min \Big(0,-\sqrt{(W^{+-}_{11} - 2W^{+-}_{22})^2 + 4|r_{62}|^2} + W^{+-}_{11} + 2W^{+-}_{22}\Big)\bigg| + \mathcal O(\lambda^4)\,.
\end{align}
\end{widetext}

Within each eigenvalue in the negativity, there is a competition between spacetime local terms and non-local terms, so that the negativity measures this overall struggle. One can see this even more explicitly if it is assumed that the density matrix components are much smaller than 1 and each system has the same parameters except for $\lambda_1 = \sqrt{2}\lambda_2$ and $L_{l_dm_{d'}}$, which yields a negativity of
\begin{align}\label{Equation:NegApprox}
&\mathcal N(\hat{\rho}_D(t)) \approx \bigg|\min \Big(0, -\frac{1}{2}\left(W^{+-}_{21}W^{+-}_{12} + |r_{46}|^2\right)\Big)\nonumber \\
&\qquad\qquad+\min \Big(0,-|r_{62}| + 2W^{+-}_{22}\Big)\bigg| + \mathcal O(\lambda^4)\,.
\end{align}

\begin{figure*}[t!]
    \includegraphics[width = 0.8 \linewidth]{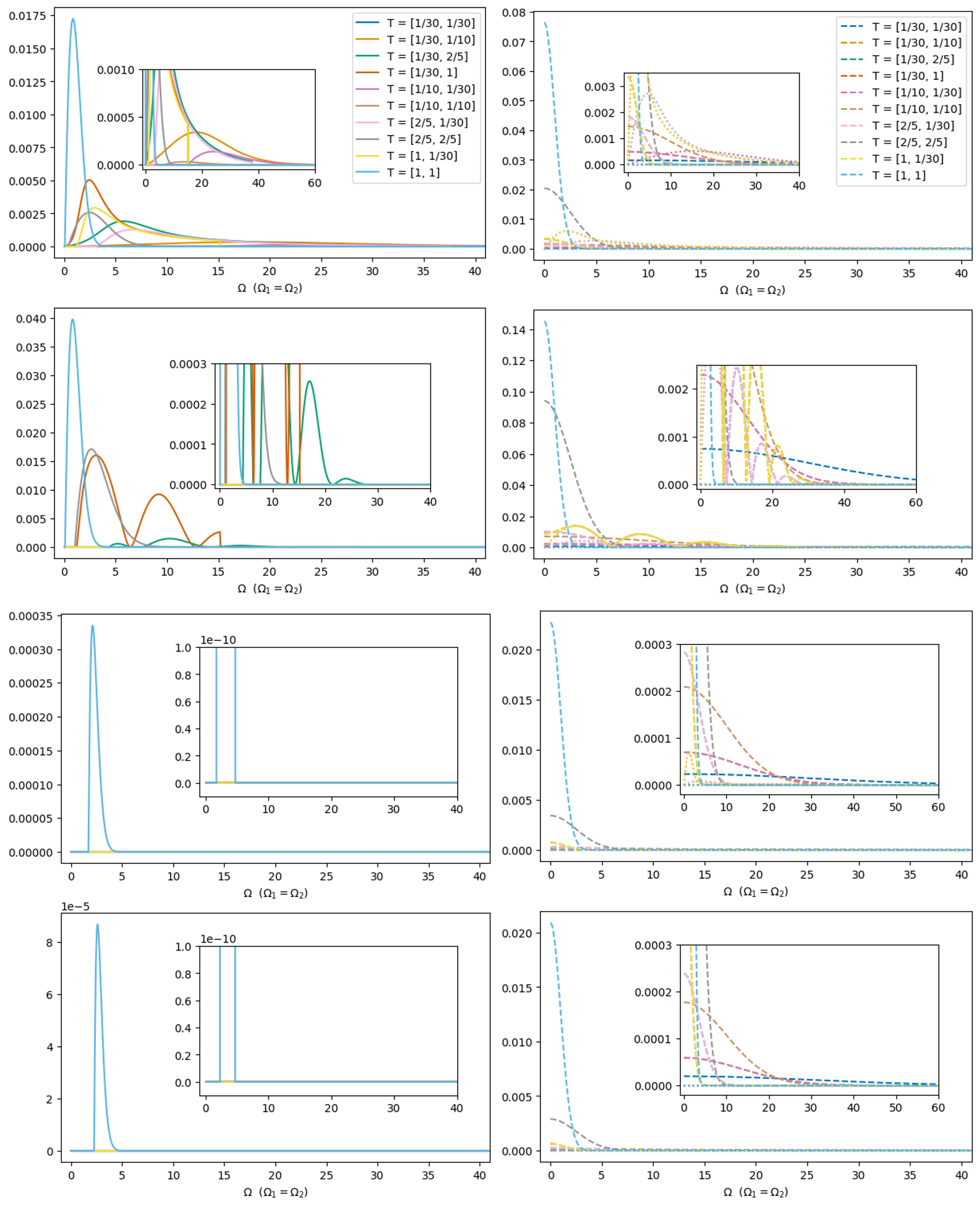}
    \caption{Plots for the qubit-qutrit harvesting scenario. The plots in the left column depict $\mathcal N(\hat{\rho}_D(t))/\lambda^2$ with solid curves, while the plots in the right column depict $|\Delta(\Lambda_d^{+}, \Lambda_{d'}^{+})|$ with dotted curves and $|H(\Lambda_d^{+}, \Lambda_{d'}^{+})|$ with dashed curves, all as functions of the energy gap $\Omega_d=\Omega_{d'}$ for different temporal smearing widths and a coupling constant of $\lambda_d = 10^{-4}$; all plots have $\bar{t}_d=0$ additionally. The first and third rows denote spatial smearing widths of $\sigma_{1_1} = 1$, while the second and third denote $\sigma_{1_1} = 0.1$. Finally, the first two rows correspond to a separation of $L_{l_dm_{d'}}=1/2$, while the last two correspond to $L_{l_dm_{d'}}=3$.}
    \label{Figure:NegQubitQutrit}
\end{figure*}

Now $\mathcal N(\hat{\rho}_D(t))/\lambda^2$,  $|\Delta(\Lambda_d^+, \Lambda^+_{d'})|$, and $|H(\Lambda_d^+, \Lambda^+_{d'})|$ are plotted in Figure 
\ref{Figure:NegQubitQutrit} for the qubit-qutrit system and compare with the harvested contextuality in the qutrit subsystem displayed in the first three columns of plots in Figure \ref{Figure:CFDManaSingleQutrit}. Notice how the amount of negativity generally increases for decreasing spatial smearing width, similarly to the amount of harvested contextuality and magic for the same temporal smearing widths. The exception to this trend are the curves corresponding to qutrit temporal smearing widths of greater than $2/5$ in the first two plots of the first column in Figure \ref{Figure:NegQubitQutrit}, which look essentially flat for smaller spatial smearing widths. This exception is most likely because sharp temporal smearing profiles coupled with smaller spatial smearing for the qutrit increase its local interaction with the field, causing the qutrit's local contribution in the second term of Eq. \eqref{Equation:NegApprox} to dominate over the cross term and the contributions of the cross terms in the first term to simultaneously dwindle, effectively isolating the qutrit and preventing the establishment of entanglement between the qutrit and qubit systems through the field. The general trend is consistent with the fact that wider local interaction regions for the two systems amplify the harvested local non-classicality and entanglement acquirable from the field. It is also apparent that, for specific measurement sets such as that corresponding to the first column of plots in Figure \ref{Figure:CFDManaSingleQutrit}, harvested contextuality is generally more available for smaller $\Omega_d$ than is harvested entanglement -- with a complete absence of negativity for specific sets of measurements (compare top left plots of Figure \ref{Figure:CFDManaSingleQutrit} and Figure \ref{Figure:NegQubitQutrit}). Furthermore, the prevalence of each resource diminishes with increasing $\Omega_d$, eventually leaving neither resource present in the system. In the intermediate range of energy gaps, the harvested contextuality appears to drop off faster than the negativity does for smaller $L_{l_dm_{d'}}$ and spatial smearing widths (compare the top row of plots in Figure \ref{Figure:CFDManaSingleQutrit} with the top left plot in Figure \ref{Figure:NegQubitQutrit}). 

Physically, the size of the energy gap determines how costly it is to change the state of the system from its initial state, which explains why both resources decay to 0 for large $\Omega_d$. The magnitude of the energy gap can also be reinterpreted in terms of the ease with which you can cross a state space boundary that delineates between non-resourceful and resourceful states according to the given resource theory. Now, because the space of contextual states and the space of entangled states have different boundaries with respect to non-contextual and separable states, respectively, it should be the case that it is easier to cross the contextual state space boundary at lower energy gap values than it is to cross the entangled state space boundary -- at least for certain sets of measurements.

The amount of harvested contextuality does not depend on the distance between systems, so for large enough $L_{l_dm_{d'}}$ only local contextuality for the qutrit remains as a quantum resource and it obviously does not influence the negativity in any way. A general trend that one can also notice in Figure \ref{Figure:NegQubitQutrit} is that for fixed qutrit temporal smearing width, the beginning of the non-zero portion of the negativity moves closer to 0 energy gap as the temporal smearing width associated with the qubit system increases. This means that one is free to fix the qutrit temporal smearing width to a value that allows for a large amount of contextuality to be harvested for small energy gap values, and then adjust the qubit temporal smearing width so that the energy gap region containing the peak value of the negativity can align with that of the qutrit's. 





\section{Conclusions and Outlook} \label{Section:Conclusions}

It was shown that a single qutrit, for which no composite structure and therefore no entanglement is possible, is nonetheless capable of harvesting non-classicality in the form of contextuality from the vacuum state of a massless scalar quantum field in the modified pentagram measurement scenario. Also somewhat surprisingly, gapless systems were seen to be able to harvest genuine contextuality -- depending on the specific measurement operators employed -- in stark contrast to the negative results for entanglement harvesting in the literature~\cite{Perche_2024}. The contextual fraction difference was introduced as a general measure of harvested contextuality, including non-local entanglement and magic, and criteria for genuine harvesting that build on earlier works was established -- extending the analysis of valid harvesting parameter regimes to individual systems. The contextual fraction difference was seen to behave similarly to other non-classicality measures such as the mana, for which the contextual fraction difference was larger in magnitude for some parameter regimes and the set of measurements corresponding to the first set of angles in Appendix \ref{Appendix:QutritContextScenario}. A qubit-qutrit setup where the qutrit can carry contextuality via the modified pentagram scenario and the combined system can have negativity was also considered. It was shown that harvested entanglement and qutrit contextuality can coexist for certain parameter regimes and influence each other, either hindering or boosting one another depending on the parameter regime.

Since the model presented here is a partially non-relativistic model that uses standard measurements for the measurement scenarios, it would be interesting to consider a fully relativistic treatment as in Reference \cite{Perche_Polo-Gómez_Torres_Martín-Martínez_2023}, to investigate how the results and conclusions here generalize. As noted in Reference \cite{Perche_Polo-Gómez_Torres_Martín-Martínez_2023}, the partially non-relativistic UDW model can be seen as a 2nd-order approximation in perturbation theory to the fully relativistic UDW model, and the amount of entanglement harvested is larger on average for the fully relativistic model, meaning there are good grounds to expect that the amount of harvested contextuality would also be larger. Employing a non-perturbative approach for the interaction should similarly yield greater amounts of harvested contextuality and would be another interesting path towards generalizing the results. 

As mentioned in the introduction, contextuality is a general resource for quantum information and computation protocols, so it would be interesting to investigate the prospect of achieving a quantum advantage in existing RQI protocols using contextuality, as well as how they might benefit from it. In particular, how contextuality would impact the capabilities of the recently proposed relativistic quantum computer model in Reference \cite{LeMaitre_Perche_Krumm_Briegel_2024} and if it could offer an extension out of the specific regimes required by that model. More realistic and noise-robust contextuality scenarios are also an important next step for eventually transitioning to experimental feasibility. 

One can also consider less pioneering extensions such as using different spacetime smearing profiles, system-field couplings, and field types to explore how the profile of the harvested contextuality changes. Moving to curved spacetimes to incorporate the gravitational field into the protocol is also possible and would be an interesting direction to see how contextuality is affected by the geometry of spacetime.

Finally, if the resource theory of contextuality is considered ~\cite{Amaral_2019}, then the idea that an initially non-contextual state could become a contextual one through interaction with a quantum field shows that this operation is clearly contextuality-generating, and is therefore not free. The extent to which one can say that the field state already has contextuality was argued at the beginning of Section \ref{Section:ContextHarvest} but is still not fully resolved. Therefore, constructing a more rigorous version of this argument geared specifically towards contextuality would be an important follow-up to this work and would also benefit the study of non-classicality in relativistic quantum field theory in general. 




\acknowledgments
The author would like to thank Adam Teixidó-Bonfill, Ahmed Shalabi, Johannes Fankhauser, and T. Rick Perche for useful discussions and comments on the paper. 
The author gratefully acknowledges support from the European Union (ERC Advanced Grant, QuantAI, No. 101055129). The views and opinions expressed in this article are however those of the author only and do not necessarily reflect those of the European Union or the European Research Council - neither the European Union nor the granting authority can be held responsible for them. For open access purposes, the author has applied a CC BY public copyright license to any author-accepted manuscript version arising from this submission.

\bibliography{context_harvest}

\appendix
\onecolumngrid
\section*{Appendices}

\section{Explicit Expressions for the Smeared Bi-Distributions, Density Matrix, and Mana} \label{Appendix:Calculations}

Recall that the spacetime smearing function used is  
\begin{align}
\Lambda^{\pm}_d(\mf x) = \Lambda_d(\mf x)e^{\pm i\Omega_d t} = \sum_{l_d} \frac{ c_{l_d}e^{-(\bm x-\bar{\bm x}_{l_d})^T\bm \Sigma^{-1}_{l_d}(\bm x-\bar{\bm x}_{l_d})-\frac{(t-\bar{t}_d)^2}{T_d^2}\pm i\Omega_d t}}{\pi^{3/2}||\tilde{c}_d||}
\end{align}
will be used where $T_d \in \mathbb R_+$ is the temporal width for system $d$, $c_{l_d} \in \mathbb R$ are the basis coefficients of each basis function corresponding to system $d$, $\bar{t}_d \in \mathbb R$ is the temporal centre for system $d$, $\bar{\bm x}_{l_d} \in \mathbb R^3$ are the spatial centre vectors corresponding to each basis function for system $d$, $||\tilde{c}_d|| = \sum_{m_d}c_{m_d}\sqrt{\text{det}\left(\bm \Sigma_{m_d}\right)}$, and $p \in \{-1, 1\}$ represents the sign associated with the monopole term. $\bm \Sigma_{l_d}$ is the 3-dimensional correlation matrix and is given by
\begin{align}
\bm \Sigma_{l_d} = \begin{pmatrix} \sigma_{1,l_d}^2 & \rho_{12, l_d} \sigma_{1,l_d}\sigma_{2,l_d} & \rho_{13, l_d}\sigma_{1,l_d}\sigma_{3,l_d} \\ \rho_{21, l_d}\sigma_{2,l_d}\sigma_{1,l_d} & \sigma_{2,l_d}^2 & \rho_{23, l_d}\sigma_{2,l_d}\sigma_{3,l_d} \\ \rho_{31, l_d}\sigma_{3,l_d}\sigma_{1,l_d} & \rho_{32, l_d}\sigma_{3,l_d}\sigma_{2,l_d} & \sigma_{3,l_d}^2\end{pmatrix}
\end{align}
where $\sigma_{i, l_d} \in \mathbb R_+$ is the spatial width corresponding to component $i$ ($i=1, 2, 3$) of each of the basis functions for system $d$ and $\rho_{ij, l_d} \in \mathbb R_+$ is the correlation between components $i$ and $j$. For further convenience, also recall the definition of the smeared version of an arbitrary bi-distribution $B(\mathsf x, \mathsf x')$:
\begin{align}
B(\Lambda^{p}_{d}, \Lambda^{q}_{d'}) = \int\mathrm{d}V_1 \int \mathrm{d}V_2 \Lambda^{p}_d(\mf x_1)\Lambda^{q}_{d'}(\mf x_2)B(\mf x_1, \mf x_2)
\end{align}
and the forward time-ordered smeared version as 
\begin{align}
B_{\Delta t}(\Lambda_d^{p}, \Lambda_{d'}^{q}) = \int\mathrm{d}V_1 \int \mathrm{d}V_2 \Lambda^{p}_d(\mf x_1)\Lambda^{q}_{d'}(\mf x_2)\theta(t_1-t_2)B(\mf x_1, \mf x_2)\,,
\end{align}
where $\theta(t-t')$ is the Heaviside step function and the backward time-ordered smeared version $B_{-\Delta t}(\Lambda_d^{p}, \Lambda_{d'}^{q}) $ is just the above with the arguments of $B(\mf x_1, \mf x_2)$ and $\theta(t_1-t_2)$ reversed. In the following calculations it will be assumed that $\bar{\bm x}_{l_d} = \bar{\bm x}_d$ and $\bm \Sigma_{l_d}$ is diagonal with each element $ \sigma_{i, l_d}$ being the same.

\subsection{Density Matrix}

First, the $\hat{\rho}_D^{(1,1)}$ term with no time-ordering is tackled, which is expanded to be
\begin{align}
\hat{\rho}_D^{(1,1)} &= \sum_{dd'}\lambda_d\lambda_{d'} \Big[\hat{J}^{(d)}_+\hat{\rho}_D(t_0) \hat{J}^{(d')}_+ W(\Lambda^{+}_{d'}, \Lambda^{+}_{d})+\hat{J}^{(d)}_+\hat{\rho}_D(t_0) \hat{J}^{(d')}_- W(\Lambda^{-}_{d'}, \Lambda^{+}_{d})\nonumber \\[1ex] 
&\qquad\qquad +\hat{J}^{(d)}_-\hat{\rho}_D(t_0) \hat{J}^{(d')}_+ W(\Lambda^{+}_{d'}, \Lambda^{-}_{d})+\hat{J}^{(d)}_-\hat{\rho}_D(t_0) \hat{J}^{(d')}_- W(\Lambda^{-}_{d'}, \Lambda^{-}_{d})\nonumber \\[1ex] 
&= \sum_{dd'}\lambda_d\lambda_{d'} \sum_{pq}\hat{J}^{(d)}_p\hat{\rho}_D(t_0) \hat{J}^{(d')}_q W(\Lambda^{q}_{d'}, \Lambda^{p}_{d})\,.
\end{align}
where $W(\Lambda^{p}_{d}, \Lambda^{q}_{d'})$ is the smeared Wightman function. Next, the $\hat{\rho}_D^{(2,0)}(t)$ term with time-ordering is expanded to be
\begin{align}
\hat{\rho}_D^{(2,0)}(t) &= -\sum_{dd'}\lambda_d\lambda_{d'} \int \mathrm{d}V_i\int \mathrm{d}V_j \Lambda_d(\mf x_i) \Lambda_{d'}(\mf x_j)\theta(t_i-t_j)\hat{\mu}_d(t_i)\hat{\mu}_{d'}(t_j)\langle\hat{\phi}(\mf x_i)\hat{\phi}(\mf x_j)\rangle_{\hat{\rho}_\phi}\hat{\rho}_D(t_0)\nonumber \\[1ex]
&=-\sum_{dd'}\lambda_d\lambda_{d'}\int \mathrm{d}V_i\int \mathrm{d}V_j  \theta(t_i-t_j)W(\mf x_i, \mf x_j)\Big(\hat{J}^{(d)}_+ \hat{J}^{(d')}_+ \Lambda^+_d(\mf x_i) \Lambda^+_{d'}(\mf x_j)\nonumber \\[1ex] 
&\quad +\hat{J}^{(d)}_+ \hat{J}^{(d')}_- \Lambda^+_d(\mf x_i) \Lambda^-_{d'}(\mf x_j)  +\hat{J}^{(d)}_- \hat{J}^{(d')}_+ \Lambda^-_d(\mf x_i) \Lambda^+_{d'}(\mf x_j) +\hat{J}^{(d)}_- \hat{J}^{(d')}_-\Lambda^-_d(\mf x_i) \Lambda^-_{d'}(\mf x_j)\Big)\hat{\rho}_D(t_0) \nonumber\\[1ex]
&=-\sum_{dd'}\lambda_d\lambda_{d'}\sum_{pq}W_{\Delta t}(\Lambda^p_d, \Lambda^q_{d'})\hat{J}^{(d)}_p \hat{J}^{(d')}_q \hat{\rho}_D(t_0)
\end{align}
where $W_{\Delta t}(\Lambda_d^{p}, \Lambda_{d'}^{q}) $ is the forward time-ordered smeared Wightman function. Now the $\hat{\rho}^{(0, 2)}(t)$ term is calculated to be
\begin{align}
\hat{\rho}_D^{(0,2)}(t) &= -\sum_{dd'}\lambda_d\lambda_{d'}\sum_{pq}W^{*}_{\Delta t}(\Lambda^p_d, \Lambda^q_{d'})\hat{\rho}_D(t_0)\left(\hat{J}^{(d')}_q\right)^{\dagger} \left(\hat{J}^{(d)}_p\right)^{\dagger}
\end{align} 
and adding the above to $\hat{\rho}_D^{(2, 0)}$ yields
\begin{align}
\hat{\rho}_D^{(2,0)}(t)+ \hat{\rho}_D^{(0,2)}(t)&= -\sum_{dd'}\lambda_d\lambda_{d'}\sum_{pq}\Big(W_{\Delta t}(\Lambda^p_d, \Lambda^q_{d'})\hat{J}^{(d)}_p\hat{J}^{(d')}_q\hat{\rho}_D(t_0)\nonumber\\[1ex]
&+W^{*}_{\Delta t}(\Lambda^p_d, \Lambda^q_{d'})\hat{\rho}_D(t_0)\left(\hat{J}^{(d')}_q\right)^{\dagger} \left(\hat{J}^{(d)}_p\right)^{\dagger} \Big)\,.
\end{align}
The final state up to second-order in $\lambda$ is then
\begin{align}
\hat{\rho}_D(t)&= \hat{\rho}_D(t_0)+\sum_{dd'}\lambda_d\lambda_{d'} \sum_{pq}\hat{J}^{(d)}_p\hat{\rho}_D(t_0) \hat{J}^{(d')}_q W(\Lambda^{q}_{d'}, \Lambda^{p}_{d})\nonumber\\[1ex]
&-\sum_{dd'}\lambda_d\lambda_{d'}\sum_{pq}\Big(W_{\Delta t}(\Lambda^p_d, \Lambda^q_{d'})\hat{J}^{(d)}_p \hat{J}^{(d')}_q\hat{\rho}_D(t_0)\nonumber\\[1ex]
&+W^{*}_{\Delta t}(\Lambda^p_d, \Lambda^q_{d'})\hat{\rho}_D(t_0)\left(\hat{J}^{(d')}_q\right)^{\dagger} \left(\hat{J}^{(d)}_p\right)^{\dagger} \Big)\,.
\end{align}
The above can be further simplified with the definition of the Feynman propagator Eq. \eqref{Equation:FeynProp} and an alternate formulation of the smeared Wightman function for $d=d'$, given by
\begin{align}\label{Equation:AltWightman}
W(\Lambda^p_d, \Lambda^q_{d}) &= W_{\Delta t}(\Lambda^p_d, \Lambda^q_{d}) + W^{*}_{\Delta t}((\Lambda^q_{d})^{*}, (\Lambda^p_{d})^{*}) = W_{-\Delta t}(\Lambda^q_d, \Lambda^p_{d}) + W^{*}_{-\Delta t}((\Lambda^p_{d})^{*}, (\Lambda^q_{d})^{*})\,.
\end{align}
Particularizing to our setting, observe that because the initial states are assumed to be in their ground state, any summand in the last term that contains at least one annihilation operator acting on the left (equivalently, a creation operator acting on the right) will be zero with the exception of the terms corresponding to $\hat{J}^{(d)}_-\hat{J}^{(d)}_+$ and its conjugate-transpose. With these simplifications, the last term becomes
\begin{align}
& -\sum_{dd'}\lambda_d\lambda_{d'}\left(W_{\Delta t}(\Lambda^+_d, \Lambda^+_{d'})\hat{J}^{(d)}_+ \hat{J}^{(d')}_+\hat{\rho}_D(t_0)+W^{*}_{\Delta t}(\Lambda^+_d, \Lambda^+_{d'})\hat{\rho}_D(t_0)\hat{J}^{(d')}_-\hat{J}^{(d)}_- \right)\nonumber\\[1ex]
& -\sum_{d}\lambda_d^2\left(W_{\Delta t}(\Lambda^-_d, \Lambda^+_{d})\hat{J}^{(d)}_- \hat{J}^{(d)}_+\hat{\rho}_D(t_0)+W^{*}_{\Delta t}(\Lambda^-_d, \Lambda^+_{d})\hat{\rho}_D(t_0)\hat{J}^{(d)}_-\hat{J}^{(d)}_+ \right) \nonumber \\[1ex]
&= -\sum_{d}\lambda_d^2\left(W_{\Delta t}(\Lambda^+_d, \Lambda^+_{d})\hat{J}^{(d)}_+ \hat{J}^{(d)}_+\hat{\rho}_D(t_0)+W^{*}_{\Delta t}(\Lambda^+_d, \Lambda^+_{d})\hat{\rho}_D(t_0)\hat{J}^{(d)}_-\hat{J}^{(d)}_- +\epsilon_d W(\Lambda^-_d,\Lambda^+_{d})\hat{\rho}_D(t_0)\right)\nonumber\\[1ex]
& -\lambda_1\lambda_2\delta_{D>1}\left(G_F(\Lambda^+_1, \Lambda^+_{2})\hat{J}^{(1)}_+ \hat{J}^{(2)}_+\hat{\rho}_D(t_0)+G^{*}_{F}(\Lambda^+_1, \Lambda^+_{2})\hat{\rho}_D(t_0)\hat{J}^{(2)}_-\hat{J}^{(1)}_- \right)
\end{align}
for $D=2$ where $\epsilon_d$ is 1 for qubit systems and 2 for qutrit systems. If $D=1$ then the last term in the above disappears. 

In the following subsections, the smeared versions of the Wightman function, Causal propagator, Hadamard function, forward/backward time-ordered Wightman function, retarded and advanced propagators, symmetric propagator, and Feynman propagator will be calculated.

\subsection{Wightman Function, Causal Propagator, and Hadamard Function}

The smeared Wightman function in 3+1 Minkowski spacetime corresponding to the same system is
\begin{align}
&W(\Lambda^{p}_{d}, \Lambda^{q}_{d}) = \sum_{l_d m_{d}} \frac{c_{l_d}c_{m_{d}}}{\pi^3||\tilde{c}_d||^2}\int^\infty_{-\infty}\mathrm{d}t_i \int^\infty_{-\infty} \mathrm{d}t_j \int \mathrm{d}\bm x_i \int \mathrm{d}\bm x_j \int \frac{\mathrm{d}\bm k}{2(2\pi)^3|\bm k|} \nonumber \\[1ex]
&\qquad\qquad\qquad\qquad\qquad \times e^{-i|\bm k|(t_i-t_j)+i\bm k\cdot (\bm x_i-\bm x_j)+ i\Omega_d(pt_i+qt_j)-\frac{(\bm x_i-\bar{\bm x}_{l_d})^2}{\sigma^2_{l_d}}-\frac{(t_i-\bar{t}_d)^2}{T_d^2}-\frac{(\bm x_j-\bar{\bm x}_{m_d})^2}{\sigma^2_{m_d}}-\frac{(t_j-\bar{t}_d)^2}{T_d^2}} \nonumber \\[1ex]
&= \sum_{l_d m_{d}} \frac{\tilde{c}_{l_d}\tilde{c}_{m_{d}}}{16\pi^3||\tilde{c}_d||^2}\int^\infty_{-\infty}\mathrm{d}t_i \int^\infty_{-\infty} \mathrm{d}t_j \int \frac{\mathrm{d}\bm k}{|\bm k|} e^{-i|\bm k|(t_i-t_j)+ i\Omega_d(pt_i+qt_j)-\frac{|\bm k|^2}{4}\beta_{l_dm_d}-\frac{(t_i-\bar{t}_d)^2}{T_d^2}-\frac{(t_j-\bar{t}_d)^2}{T_d^2}} \nonumber \\[1ex]
&= \sum_{l_d m_{d}} \frac{T_d^2\tilde{c}_{l_d}\tilde{c}_{m_{d}}e^{-\frac{1}{2}(T_d\Omega_d)^2+i(p+q)\Omega_d\bar{t}_d}}{4\pi||\tilde{c}_d||^2} \int^\infty_0 \mathrm{d}|\bm k||\bm k| e^{-\frac{|\bm k|^2}{4}(\beta_{l_dm_d}+2T_d^2)+\frac{|\bm k|}{2}T_d^2\Omega_d(p-q)} \nonumber \\[1ex]
&= \sum_{l_d m_{d}} \frac{T_d^2\tilde{c}_{l_d}\tilde{c}_{m_{d}}e^{-\frac{1}{2}(T_d\Omega_d)^2+i(p+q)\Omega_d\bar{t}_d}}{2\pi||\tilde{c}_d||^2(\beta_{l_dm_d}+2T_d^2)^{\frac{3}{2}}} \left(\sqrt{\beta_{l_dm_d}+2T_d^2}+ \frac{\sqrt{\pi}}{2}T_d^2 \Omega_d(p-q)e^{\frac{((p-q)\Omega_dT_d^2)^2}{4(\beta_{l_dm_d}+2T_d^2)}}\left(1+\text{erf}\left(\frac{T_d^2\Omega_d(p-q)}{2\sqrt{\beta_{l_dm_d}+2T_d^2}}\right)\right)\right)\,.
\end{align}
with $\beta_{l_dm_{d'}} = \sigma_{l_d}^{2}+\sigma_{m_{d'}}^{2}$.

For different systems, the smeared Wightman function is
\begin{align}
&W(\Lambda^{p}_{d}, \Lambda^{q}_{d'}) = \sum_{l_d m_{d'}}  \frac{c_{l_d}c_{m_{d'}}}{\pi^3T_dT_{d'}||\tilde{c}_d||||\tilde{c}_{d'}||}\int^\infty_{-\infty}\mathrm{d}t_i \int^\infty_{-\infty} \mathrm{d}t_j \int \mathrm{d}\bm x_i \int \mathrm{d} \bm x_j \int \frac{\mathrm{d}\bm k}{2(2\pi)^3|\bm k|} \nonumber\\[1ex]
&\qquad\qquad\qquad\qquad \times e^{-i|\bm k|(t_i-t_j)+i\bm k\cdot (\bm x_i-\bm x_j)+ i(p\Omega_dt_i+q\Omega_{d'}t_j)-\frac{(\bm x_i-\bar{\bm x}_{l_d})^2}{\sigma^2_{l_d}}-\frac{(t_i-\bar{t}_d)^2}{T_d^2}-\frac{(\bm x_j-\bar{\bm x}_{m_{d'}})^2}{\sigma^2_{m_{d'}}}-\frac{(t_j-\bar{t}_{d'})^2}{T_{d'}^2}} \nonumber \\[1ex]
&= \sum_{l_d m_{d'}} \frac{\tilde{c}_{l_d}\tilde{c}_{m_{d'}}}{16\pi^3||\tilde{c}_d||||\tilde{c}_{d'}||}\int^\infty_{-\infty}\mathrm{d}t_i \int^\infty_{-\infty} \mathrm{d}t_j \int \frac{\mathrm{d}\bm k}{|\bm k|} \nonumber\\[1ex]
&\qquad\qquad\qquad\qquad \times e^{-i|\bm k|(t_i-t_j)+ i(p\Omega_dt_i+q\Omega_{d'}t_j)-\frac{|\bm k|^2}{4}\beta_{l_dm_{d'}}+i\bm k \cdot(\bar{\bm x}_{l_d}-\bar{\bm x}_{m_{d'}})-\frac{(t_i-\bar{t}_d)^2}{T_d^2}-\frac{(t_j-\bar{t}_{d'})^2}{T_{d'}^2}} \nonumber \\[1ex]
&= \sum_{l_d m_{d'}} \frac{\pi\tilde{c}_{l_d}\tilde{c}_{m_{d'}}}{8iL_{l_dm_{d'}}\pi^3||\tilde{c}_d||||\tilde{c}_{d'}||}\int^\infty_{-\infty}\mathrm{d}t_i \int^\infty_{-\infty} \mathrm{d}t_j \int^\infty_0 \mathrm{d}|\bm k| \nonumber\\[1ex]
&\qquad\qquad\qquad\qquad \times e^{-i|\bm k|(t_i-t_j)+ i(p\Omega_dt_i+q\Omega_{d'}t_j)-\frac{|\bm k|^2}{4}\beta_{l_dm_{d'}}-\frac{(t_i-\bar{t}_d)^2}{T_d^2}-\frac{(t_j-\bar{t}_{d'})^2}{T_{d'}^2}}(e^{i|\bm k|L_{l_dm_{d'}}}-e^{-i|\bm k|L_{l_dm_{d'}}}) \nonumber \\[1ex]
&= \sum_{l_d m_{d'}} \frac{T_dT_{d'}\tilde{c}_{l_d}\tilde{c}_{m_{d'}}e^{-\frac{1}{4}((T_d\Omega_d)^2+(T_{d'}\Omega_{d'})^2)+ i(p\Omega_d \bar{t}_d+q\Omega_{d'}\bar{t}_{d'})}}{8iL_{l_dm_{d'}}\pi||\tilde{c}_d||||\tilde{c}_{d'}||} \int^\infty_0 \mathrm{d}|\bm k| \nonumber\\[1ex]
&\qquad\qquad\qquad \qquad\times e^{\frac{1}{4}(-|\bm k|^2(\beta_{l_dm_{d'}}+T_d^2+T_{d'}^2)+2|\bm k|(p\Omega_dT_d^2-q\Omega_{d'}T_{d'}^2-2i(\bar{t}_d-\bar{t}_{d'}))}(e^{i|\bm k|L_{l_dm_{d'}}}-e^{-i|\bm k|L_{l_dm_{d'}}}) \nonumber \\[1ex]
&= \sum_{l_d m_{d'}} \frac{T_dT_{d'}\tilde{c}_{l_d}\tilde{c}_{m_{d'}}e^{-\frac{1}{4}((T_d\Omega_d)^2+(T_{d'}\Omega_{d'})^2)+ i(p\Omega_d \bar{t}_d+q\Omega_{d'}\bar{t}_{d'})}}{8iL_{l_dm_{d'}}\sqrt{\pi}||\tilde{c}_d||||\tilde{c}_{d'}||\sqrt{\beta_{l_dm_{d'}}+T_d^2+T_{d'}^2}}\Bigg[\nonumber \\[1ex]
&e^{\frac{(p \Omega_dT_d^2-q\Omega_{d'}T_{d'}^2+2i(\bar{t}_{d'}-\bar{t}_{d}+L_{l_dm_{d'}}))^2}{4(\beta_{l_dm_{d'}}+T_d^2+T_{d'}^2)}}\left(1+i\text{erfi}\left(\frac{2(\bar{t}_{d'}-\bar{t}_d+L_{l_dm_{d'}})-i(p \Omega_dT_d^2-q\Omega_{d'}T_{d'}^2)}{2\sqrt{\beta_{l_dm_{d'}}+T_d^2+T_{d'}^2}}\right)\right)\nonumber \\[1ex]
&-e^{\frac{(p \Omega_dT_d^2-q\Omega_{d'}T_{d'}^2+2i(\bar{t}_{d'}-\bar{t}_{d}-L_{l_dm_{d'}}))^2}{4(\beta_{l_dm_{d'}}+T_d^2+T_{d'}^2)}}\left(1+i\text{erfi}\left(\frac{2(\bar{t}_{d'}-\bar{t}_d-L_{l_dm_{d'}})-i(p \Omega_dT_d^2-q\Omega_{d'}T_{d'}^2)}{2\sqrt{\beta_{l_dm_{d'}}+T_d^2+T_{d'}^2}}\right)\right)\Bigg] 
\end{align}
with $L_{l_dm_{d'}} = ||\bar{\bm x}_d-\bar{\bm x}_{d'}||$. The smeared Hadamard term can then be calculated from its definition as
\begin{align}
H(\Lambda^p_d, \Lambda^q_{d'}) = W(\Lambda^p_d, \Lambda^q_{d'}) + W(\Lambda^q_{d'}, \Lambda^p_{d})
\end{align}
and the smeared causal propagator term from its definition as
\begin{align}
iE(\Lambda^p_d, \Lambda^q_{d'}) = W(\Lambda^p_d, \Lambda^q_{d'}) - W(\Lambda^q_{d'}, \Lambda^p_{d})\,.
\end{align}
Swapping the argument of the Wightman function amounts to reversing the order of $p$ and $q$ in the same system case, and attaching minus signs to the erfi terms and swapping the signs of $L_{l_dm_{d'}}$ between the two terms in the different systems case. Thus
\begin{align} \label{Equation:Hadamard}
H(\Lambda^p_d, \Lambda^q_{d'}) &= \sum_{l_d m_{d}} \frac{\delta_{dd'}T_d^2\tilde{c}_{l_d}\tilde{c}_{m_{d}}e^{-\frac{1}{2}(T_d\Omega_d)^2+i(p+q)\Omega_d\bar{t}_d}}{\pi||\tilde{c}_d||^2(\beta_{l_dm_d}+2T_d^2)^{\frac{3}{2}}}\left(\sqrt{\beta_{l_dm_d}+2T_d^2}+ \frac{\sqrt{\pi}}{2}T_d^2 \Omega_d(p-q)e^{\frac{((p-q)\Omega_dT_d^2)^2}{4(\beta_{l_dm_d}+2T_d^2)}}\text{erf}\left(\frac{T_d^2\Omega_d(p-q)}{2\sqrt{\beta_{l_dm_d}+2T_d^2}}\right)\right)\nonumber\\[1ex]
&+ \sum_{l_d m_{d'}} \frac{(1-\delta_{dd'})T_dT_{d'}\tilde{c}_{l_d}\tilde{c}_{m_{d'}}e^{-\frac{1}{4}((T_d\Omega_d)^2+(T_{d'}\Omega_{d'})^2)+ i(p\Omega_d \bar{t}_d+q\Omega_{d'}\bar{t}_{d'})}}{4L_{l_dm_{d'}}\sqrt{\pi}||\tilde{c}_d||||\tilde{c}_{d'}||\sqrt{\beta_{l_dm_{d'}}+T_d^2+T_{d'}^2}}\Bigg[\nonumber \\[1ex]
&e^{\frac{(p \Omega_dT_d^2-q\Omega_{d'}T_{d'}^2+2i(\bar{t}_{d'}-\bar{t}_{d}+L_{l_dm_{d'}}))^2}{4(\beta_{l_dm_{d'}}+T_d^2+T_{d'}^2)}}\text{erfi}\left(\frac{2(\bar{t}_{d'}-\bar{t}_d+L_{l_dm_{d'}})-i(p \Omega_dT_d^2-q\Omega_{d'}T_{d'}^2)}{2\sqrt{\beta_{l_dm_{d'}}+T_d^2+T_{d'}^2}}\right)\nonumber \\[1ex]
&-e^{\frac{(p \Omega_dT_d^2-q\Omega_{d'}T_{d'}^2+2i(\bar{t}_{d'}-\bar{t}_{d}-L_{l_dm_{d'}}))^2}{4(\beta_{l_dm_{d'}}+T_d^2+T_{d'}^2)}}\text{erfi}\left(\frac{2(\bar{t}_{d'}-\bar{t}_d-L_{l_dm_{d'}})-i(p \Omega_dT_d^2-q\Omega_{d'}T_{d'}^2)}{2\sqrt{\beta_{l_dm_{d'}}+T_d^2+T_{d'}^2}}\right)\Bigg] 
\end{align}
and 
\begin{align}
iE(\Lambda^p_d, \Lambda^q_{d'}) &= \sum_{l_d m_{d}} \frac{\delta_{dd'}T_d^4\Omega_d(p-q)\tilde{c}_{l_d}\tilde{c}_{m_{d}}e^{-\frac{1}{2}(T_d\Omega_d)^2+i(p+q)\Omega_d\bar{t}_d}}{2\sqrt{\pi}||\tilde{c}_d||^2(\beta_{l_dm_d}+2T_d^2)^{\frac{3}{2}}} e^{\frac{((p-q)\Omega_dT_d^2)^2}{4(\beta_{l_dm_d}+2T_d^2)}}\nonumber\\[1ex]
&+ \sum_{l_d m_{d'}} \frac{(1-\delta_{dd'})T_dT_{d'}\tilde{c}_{l_d}\tilde{c}_{m_{d'}}e^{-\frac{1}{4}((T_d\Omega_d)^2+(T_{d'}\Omega_{d'})^2)+ i(p\Omega_d \bar{t}_d+q\Omega_{d'}\bar{t}_{d'})}}{4iL_{l_dm_{d'}}\sqrt{\pi}||\tilde{c}_d||||\tilde{c}_{d'}||\sqrt{\beta_{l_dm_{d'}}+T_d^2+T_{d'}^2}}\Bigg[\nonumber\\[1ex]
&e^{\frac{(p \Omega_dT_d^2-q\Omega_{d'}T_{d'}^2+2i(\bar{t}_{d'}-\bar{t}_{d}+L_{l_dm_{d'}}))^2}{4(\beta_{l_dm_{d'}}+T_d^2+T_{d'}^2)}}-e^{\frac{(p \Omega_dT_d^2-q\Omega_{d'}T_{d'}^2+2i(\bar{t}_{d'}-\bar{t}_{d}-L_{l_dm_{d'}}))^2}{4(\beta_{l_dm_{d'}}+T_d^2+T_{d'}^2)}}\Bigg] 
\end{align}

\subsection{Forward and Backward Time-Ordered Wightman Functions}

The forward time-ordered smeared Wightman function for $t_i>t_j$ corresponding to the same system is then
\begin{align}
&\sum_{l_d m_{d}}  \frac{c_{l_d}c_{m_{d}}}{\pi^3||\tilde{c}_d||^2}\int^\infty_{-\infty}\mathrm{d}t_i \int^\infty_{-\infty} \mathrm{d}t_j \theta(t_i-t_j)\int \mathrm{d}\bm x_i \int \mathrm{d}\bm x_j \int \frac{\mathrm{d}\bm k}{2(2\pi)^3|\bm k|} \nonumber\\[1ex]
&\qquad\qquad\qquad\qquad \times e^{-i|\bm k|(t_i-t_j)+\bm k\cdot (\bm x_i-\bm x_j)+ i\Omega_d(p t_i+q t_j)-\frac{(\bm x_i-\bar{\bm x}_{l_d})^2}{\sigma^2_{l_d}}-\frac{(t_i-\bar{t}_d)^2}{T_d^2}-\frac{(\bm x_j-\bar{\bm x}_{m_d})^2}{\sigma^2_{m_d}}-\frac{(t_j-\bar{t}_d)^2}{T_d^2}} \nonumber \\[1ex]
&= \sum_{l_d m_{d}}  \frac{\pi \tilde{c}_{l_d}\tilde{c}_{m_{d}}}{4\pi^3||\tilde{c}_d||^2}\int^\infty_{-\infty}\mathrm{d}t_i \int^\infty_{-\infty} \mathrm{d}t_j \theta(t_i-t_j)\int^\infty_0 \mathrm{d}|\bm k||\bm k| e^{-i|\bm k|(t_i-t_j)+ i\Omega_d(p t_i+q t_j)-\frac{|\bm k|^2}{4}\beta_{l_dm_d}-\frac{(t_i-\bar{t}_d)^2}{T_d^2}-\frac{(t_j-\bar{t}_d)^2}{T_d^2}} \nonumber \\[1ex]
\end{align}
now focus on the time integrals and make the substitutions $t_i-t_j=x, t_i+t_j=y$:
\begin{align}
&\frac{1}{2}\int^\infty_{-\infty}\mathrm{d}y \int^\infty_{-\infty} \mathrm{d}x \theta(x)e^{-i|\bm k|x+i\Omega_d(p (x+y)+q (y-x))/2 -\frac{1}{T_d^2}(((x+y)/2-\bar{t}_d)^2+((y-x)/2-\bar{t}_d)^2)} \nonumber \\[1ex]
&= \frac{1}{2}\int^\infty_{-\infty}\mathrm{d}y \int^\infty_{0} \mathrm{d}x e^{-i|\bm k|x+i\Omega_d(p (x+y)+q (y-x))/2 -\frac{1}{T_d^2}(((x+y)/2-\bar{t}_d)^2+((y-x)/2-\bar{t}_d)^2)} \nonumber \\[1ex]
&= \frac{\pi T_d^2}{2}e^{-\frac{1}{2}(|\bm k|^2 - |\bm k|\Omega_d(p-q) + \Omega_d^2)T_d^2 + i\Omega_d(p+q)\bar{t}_d}\left(1 - i\text{erfi}\left(\frac{(2|\bm k| - \Omega_d(p-q))T_d}{2\sqrt{2}}\right)\right)
\end{align}
which leads to
\begin{align}
&\sum_{l_d m_{d}} \frac{T_d^2\tilde{c}_{l_d}\tilde{c}_{m_{d}}e^{i\Omega_d(p+q)\bar{t}_d}}{8\pi||\tilde{c}_d||^2} \int^\infty_0 \mathrm{d}|\bm k||\bm k| e^{-\frac{|\bm k|^2}{4}\beta_{l_dm_d}-\frac{1}{2}(|\bm k|^2 - |\bm k|\Omega_d(p-q) + \Omega_d^2)T_d^2}\left(1 - i\text{erfi}\left(\frac{(2|\bm k| - \Omega_d(p-q))T_d}{2\sqrt{2}}\right)\right)\nonumber \\[1ex]
&= \sum_{l_d m_{d}} \frac{T_d^2\tilde{c}_{l_d}\tilde{c}_{m_{d}}}{4\pi||\tilde{c}_d||^2(\beta_{l_dm_d}+2T_d^2)} e^{\pm2i\Omega_d\bar{t}_d-\frac{1}{2}(\Omega_dT_d)^2}\left(1-\frac{iT_d\sqrt{2}}{\sqrt{\beta_{l_dm_d}}}\right)
\end{align}
for same-sign smearing functions, and
\begin{align}
&\sum_{l_d m_{d}} \frac{T_d^2\tilde{c}_{l_d}\tilde{c}_{m_{d}}}{8\pi||\tilde{c}_d||^2}\Bigg[\frac{2e^{-\frac{1}{2}(\Omega_dT_d)^2}}{(\beta_{l_dm_{d}}+2T_d^2)^{3/2}}\left(\sqrt{\beta_{l_dm_{d}}+2T_d^2}\pm \sqrt{\pi}\Omega_dT_d^2e^{\frac{\Omega_d^2T_d^4}{\beta_{l_dm_{d}}+2T_d^2}}\left(1+\text{erf}\left(\frac{\pm \Omega_dT_d^2}{\sqrt{\beta_{l_dm_{d}}+2T_d^2}}\right)\right)\right) \nonumber\\[1ex]
&-i\int^\infty_0 \mathrm{d}|\bm k||\bm k| e^{-\frac{|\bm k|^2}{4}\beta_{l_dm_d}-\frac{1}{2}(|\bm k|- \pm \Omega_d)^2T_d^2}\text{erfi}\left(\frac{(|\bm k| -\pm \Omega_d)T_d}{\sqrt{2}}\right)\Bigg]
\end{align}
for different signs. The backward time-ordered smeared Wightman function for $t_j>t_i$ corresponding to the same system is
\begin{align}
&\sum_{l_d m_{d}} \frac{c_{l_d}c_{m_{d}}}{\pi^3||\tilde{c}_d||^2}\int^\infty_{-\infty}\mathrm{d}t_i \int^\infty_{-\infty} \mathrm{d}t_j \Theta(t_j-t_i)\int \mathrm{d}\bm x_i \int \mathrm{d}\bm x_j \int \frac{\mathrm{d}\bm k}{2(2\pi)^3|\bm k|} \nonumber\\[1ex]
&\qquad\qquad\qquad\qquad \times e^{i|\bm k|(t_i-t_j)-\bm k\cdot (\bm x_i- \bm x_j)+ i\Omega_d(p t_i+q t_j)-\frac{(\bm x_i-\bar{\bm x}_{l_d})^2}{\sigma^2_{l_d}}-\frac{(t_i-\bar{t}_d)^2}{T_d^2}-\frac{(\bm x_j-\bar{\bm x}_{m_d})^2}{\sigma^2_{m_d}}-\frac{(t_j-\bar{t}_d)^2}{T_d^2}} \nonumber \\[1ex]
&= \sum_{l_d m_{d}} \frac{\pi \tilde{c}_{l_d}\tilde{c}_{m_{d}}}{4\pi^3||\tilde{c}_d||^2}\int^\infty_{-\infty}\mathrm{d}t_i \int^\infty_{-\infty} \mathrm{d}t_j \Theta(t_j-t_i)\int^\infty_0 \mathrm{d}|\bm k||\bm k| e^{i|\bm k|(t_i-t_j)+ i\Omega_d(p t_i+q t_j)-\frac{|\bm k|^2}{4}\beta_{l_dm_d}-\frac{(t_i-\bar{t}_d)^2}{T_d^2}-\frac{(t_j-\bar{t}_d)^2}{T^2}}
\end{align}
now focus on the time integrals and make the substitutions $t_i-t_j=x, t_i+t_j=y$:
\begin{align}
&\frac{1}{2}\int^\infty_{-\infty}\mathrm{d}y \int^\infty_{-\infty} \mathrm{d}x \Theta(-x)e^{i|\bm k|x+i\Omega_d(p (x+y)+q (y-x))/2 -\frac{1}{T_d^2}(((x+y)/2-\bar{t}_d)^2+((y-x)/2-\bar{t}_d)^2)} \nonumber \\[1ex]
&= \frac{1}{2}\int^\infty_{-\infty}\mathrm{d}y \int^0_{-\infty} \mathrm{d}x e^{i|\bm k|x+i\Omega_d(p (x+y)+q (y-x))/2 -\frac{1}{T_d^2}(((x+y)/2-\bar{t}_d)^2+((y-x)/2-\bar{t}_d)^2)} \nonumber \\[1ex]
&= \frac{\pi T_d^2}{2}e^{-\frac{1}{2}(|\bm k|^2 + |\bm k|\Omega_d(p-q) + \Omega_d^2)T_d^2 + i\Omega_d(p+q)\bar{t}_d}\left(1 - i\text{erfi}\left(\frac{(2|\bm k| + \Omega_d(p-q))T_d}{2\sqrt{2}}\right)\right)
\end{align}
which leads to
\begin{align}
&\sum_{l_d m_{d}} \frac{T_d^2\tilde{c}_{l_d}\tilde{c}_{m_{d}}e^{i\Omega_d(p+q)\bar{t}_d}}{8\pi||\tilde{c}_d||^2} \int^\infty_0 \mathrm{d}|\bm k||\bm k| e^{-\frac{|\bm k|^2}{4}\beta_{l_dm_d}-\frac{1}{2}(|\bm k|^2 + |\bm k|\Omega_d(p-q) + \Omega_d^2)T_d^2}\left(1 - i\text{erfi}\left(\frac{(2|\bm k| + \Omega_d(p-q))T_d}{2\sqrt{2}}\right)\right)\nonumber \\[1ex]
&= \sum_{l_d m_{d}} \frac{T_d^2\tilde{c}_{l_d}\tilde{c}_{m_{d}}}{4\pi||\tilde{c}_d||^2(\beta_{l_dm_d}+2T_d^2)} e^{\pm2i\Omega_d\bar{t}_d-\frac{1}{2}(\Omega_dT_d)^2}\left(1-\frac{iT_d\sqrt{2}}{\sqrt{\beta_{l_dm_d}}}\right)
\end{align}
for same-sign smearing functions, and
\begin{align}
&\sum_{l_d m_{d}} \frac{T_d^2\tilde{c}_{l_d}\tilde{c}_{m_{d}}}{8\pi||\tilde{c}_d||^2}\Bigg[\frac{2e^{-\frac{1}{2}(\Omega_dT_d)^2}}{(\beta_{l_dm_{d}}+2T_d^2)^{3/2}}\left(\sqrt{\beta_{l_dm_{d}}+2T_d^2}-\pm \sqrt{\pi}\Omega_dT_d^2e^{\frac{\Omega_d^2T_d^4}{\beta_{l_dm_{d}}+2T_d^2}}\text{erfc}\left(\frac{\pm \Omega_dT_d^2}{\sqrt{\beta_{l_dm_{d}}+2T_d^2}}\right)\right) \nonumber\\[1ex]
&-i\int^\infty_0 \mathrm{d}|\bm k||\bm k| e^{-\frac{|\bm k|^2}{4}\beta_{l_dm_d}-\frac{1}{2}(|\bm k| \pm \Omega_d)^2T_d^2}\text{erfi}\left(\frac{(|\bm k| \pm \Omega_d)T_d}{\sqrt{2}}\right)\Bigg]
\end{align}
for different signs. For different systems and $t_i>t_j$, the forward time-ordered smeared Wightman function is
\begin{align}
&\sum_{l_d m_{d'}}\frac{ c_{l_d}c_{m_{d'}}}{\pi^3||\tilde{c}_d||||\tilde{c}_{d'}||}\int^\infty_{-\infty}\mathrm{d}t_i \int^\infty_{-\infty} \mathrm{d}t_j \Theta(t_i-t_j)\int \mathrm{d}\bm x_i \int \mathrm{d}\bm x_j \int \frac{\mathrm{d}\bm k}{2(2\pi)^3|\bm k|}\nonumber\\[1ex]
&\qquad\qquad\qquad\qquad \times e^{-i|\bm k|(t_i-t_j)+\bm k\cdot (\bm x_i-\bm x_j)+ i(p\Omega_d t_i+q \Omega_{d'}t_j)-\frac{(\bm x_i-\bar{\bm x}_{l_d})^2}{\sigma^2_{l_d}}-\frac{(t_i-\bar{t}_d)^2}{T_d^2}-\frac{(\bm x_j-\bar{\bm x}_{m_{d'}})^2}{\sigma^2_{m_{d'}}}-\frac{(t_j-\bar{t}_{d'})^2}{T_{d'}^2}} \nonumber \\[1ex]
&= \sum_{l_d m_{d'}} \frac{\tilde{c}_{l_d}\tilde{c}_{m_{d'}}}{16\pi^3||\tilde{c}_d||||\tilde{c}_{d'}||}\int^\infty_{-\infty}\mathrm{d}t_i \int^\infty_{-\infty} \mathrm{d}t_j \Theta(t_i-t_j)\int \frac{\mathrm{d}\bm k}{|\bm k|} \nonumber\\[1ex]
&\qquad\qquad\qquad\qquad \times e^{-i|\bm k|(t_i-t_j)+i\bm k \cdot(\bar{\bm x}_{l_d}-\bar{\bm x}_{m_{d'}}) + i(p\Omega_d t_i+q \Omega_{d'}t_j)-\frac{|\bm k|^2}{4}\beta_{l_dm_{d'}}-\frac{(t_i-\bar{t}_d)^2}{T_d^2}-\frac{(t_j-\bar{t}_{d'})^2}{T_{d'}^2}} \nonumber \\[1ex]
&= \sum_{l_d m_{d'}} \frac{\tilde{c}_{l_d}\tilde{c}_{m_{d'}}}{8iL_{l_dm_{d'}}||\tilde{c}_d||||\tilde{c}_{d'}||}\int^\infty_{-\infty}\mathrm{d}t_i \int^\infty_{-\infty} \mathrm{d}t_j \Theta(t_i-t_j)\int^\infty_0\mathrm{d}|\bm k| \left(e^{i|\bm k|L_{l_dm_{d'}}}-e^{-i|\bm k|L_{l_dm_{d'}}}\right)\nonumber \\[1ex]
&\qquad\qquad\qquad\qquad\qquad\qquad\qquad\qquad \times e^{-i|\bm k|(t_i-t_j) + i(p\Omega_d t_i+q \Omega_{d'}t_j)-\frac{|\bm k|^2}{4}\beta_{l_dm_{d'}}-\frac{(t_i-\bar{t}_d)^2}{T_d^2}-\frac{(t_j-\bar{t}_{d'})^2}{T_{d'}^2}}
\end{align}
where the $\bm k$ integrals have been reduced  to one integral by choosing the direction $\bm k \cdot (\bar{x}_d-\bar{x}_{d'}) = |\bm k||\bar{x}_d-\bar{x}_{d'}|\cos{(k_\theta)} = |\bm k|L_{l_dm_{d'}}\cos{(k_\theta)}$:
\begin{align}
\int \frac{\mathrm{d}\bm k}{|\bm k|}e^{i|\bm k| L_{l_dm_{d'}}\cos{(k_\theta)}-i|\bm k|(t_i-t_j) -\frac{|\bm k|^2}{4}\beta_{l_dm_{d'}}} &= \int^{2 \pi}_0\mathrm{d}k_\phi \int^1_{-1} \mathrm{d}(\cos{(k_\theta)}) \int^\infty_0 \mathrm{d}|\bm k|  |\bm k| e^{i|\bm k| L_{l_dm_{d'}}\cos{(k_\theta)}-i|\bm k|(t_i-t_j) -\frac{|\bm k|^2}{4}\beta_{l_dm_{d'}}} \nonumber \\[1ex]
&= \int^\infty_0 \mathrm{d}|\bm k|\frac{2 \pi}{iL_{l_dm_{d'}}} \left(e^{i|\bm k|L_{l_dm_{d'}}}-e^{-i|\bm k|L_{l_dm_{d'}}}\right)e^{-i|\bm k|(t_i-t_j) -\frac{|\bm k|^2}{4}\beta_{l_dm_{d'}}} \nonumber
\end{align}
Now focus on the time integrals again and make the substitutions $t_i-t_j=x, t_i+t_j=y$, then
\begin{align}
&\frac{1}{2}\int^\infty_{-\infty}\mathrm{d}y \int^\infty_{-\infty} \mathrm{d}x \Theta(x)e^{-i|\bm k|x+ \frac{i}{2}(p (x+y)\Omega_d+q (y-x)\Omega_{d'}) -\frac{1}{T_d^2}((x+y)/2-\bar{t}_d)^2-\frac{1}{T_{d'}^2}((y-x)/2-\bar{t}_{d'})^2} \nonumber \\[1ex]
&= \frac{1}{2}\int^\infty_{-\infty}\mathrm{d}y \int^\infty_{0} \mathrm{d}x e^{-i|\bm k|x+ \frac{i}{2}(p (x+y)\Omega_d+q (y-x)\Omega_{d'}) -\frac{1}{T_d^2}((x+y)/2-\bar{t}_d)^2-\frac{1}{T_{d'}^2}((y-x)/2-\bar{t}_{d'})^2} \nonumber \\[1ex]
&= \frac{\pi T_d^2T_{d'}^2}{\sqrt{2(T_d^4+T_{d'}^4)}}e^{\frac{T_d^2T_{d'}^2}{2(T_d^4+T_{d'}^4)}\left(-|\bm k|^2(T_d^2+T_{d'}^2)+|\bm k|(2T_d^2(p\Omega_d-q\Omega_{d'})-4i(\bar{t}_d-\bar{t}_{d'}))\right)}\nonumber\\[1ex]
&\qquad\quad\times e^{\frac{1}{4(T_d^4+T_{d'}^4)(T_d^2+T_{d'}^2)}\left(4(\bar{t}_d-\bar{t}_{d'})^2((T_d^2+T_{d'}^2)^2-T_d^2T_{d'}^2)-(3T_d^4+T_{d'}^4)T_d^2T_{d'}^2(p\Omega_d-q\Omega_{d'})^2+4(p\Omega_d-q\Omega_{d'})(T_{d'}^2\bar{t}_d(3T_d^4+T_{d'}^4)+T_d^2\bar{t}_{d'}(T_d^4-T_{d'}^4))\right)}\nonumber\\[1ex]
&\qquad\quad\times \left(1+\text{erf}\left(\frac{(2(\bar{t}_d-\bar{t}_{d'})-i|\bm k|(T_d^2+T_{d'}^2)+iT_d^2(p\Omega_d-q\Omega_{d'}))T_dT_{d'}}{\sqrt{2(T_d^4+T_{d'}^4)(T_d^2+T_{d'}^2)}}\right)\right) 
\end{align}
which leads to
\begin{align}
&\sum_{l_d m_{d'}} \frac{(T_dT_{d'})^2\tilde{c}_{l_d}\tilde{c}_{m_{d'}}}{8iL_{l_dm_{d'}}\pi||\tilde{c}_d||||\tilde{c}_{d'}||\sqrt{2(T_d^4+T_{d'}^4)}} \int^\infty_0 \mathrm{d}|\bm k| e^{\frac{T_d^2T_{d'}^2}{2(T_d^4+T_{d'}^4)}\left(-|\bm k|^2(T_d^2+T_{d'}^2)+|\bm k|(2T_d^2(p\Omega_d-q\Omega_{d'})-4i(\bar{t}_d-\bar{t}_{d'}))\right)}\nonumber\\[1ex]
&\times  e^{-\frac{|\bm k|^2}{4}\beta_{l_dm_{d'}}}(e^{i|\bm k|L_{l_dm_{d'}}}-e^{-i|\bm k|L_{l_dm_{d'}}})\left(1+\text{erf}\left(\frac{(2(\bar{t}_d-\bar{t}_{d'})-i|\bm k|(T_d^2+T_{d'}^2)+iT_d^2(p\Omega_d-q\Omega_{d'}))T_dT_{d'}}{\sqrt{2(T_d^4+T_{d'}^4)(T_d^2+T_{d'}^2)}}\right)\right)\nonumber\\[1ex]
&\times e^{\frac{1}{4(T_d^4+T_{d'}^4)(T_d^2+T_{d'}^2)}\left(4(\bar{t}_d-\bar{t}_{d'})^2((T_d^2+T_{d'}^2)^2-T_d^2T_{d'}^2)-(3T_d^4+T_{d'}^4)T_d^2T_{d'}^2(p\Omega_d-q\Omega_{d'})^2+4(p\Omega_d-q\Omega_{d'})(T_{d'}^2\bar{t}_d(3T_d^4+T_{d'}^4)+T_d^2\bar{t}_{d'}(T_d^4-T_{d'}^4))\right)}\,.
\end{align}
Now for the case when $t_j>t_i$, the backward time-ordered smeared Wightman function corresponding to different systems is
\begin{align}
&\sum_{l_d m_{d'}} \frac{c_{l_d}c_{m_{d'}}}{\pi^3||\tilde{c}_d||||\tilde{c}_{d'}||}\int^\infty_{-\infty}\mathrm{d}t_i \int^\infty_{-\infty} \mathrm{d}t_j \Theta(t_j-t_i)\int \mathrm{d}\bm x_i \int \mathrm{d}\bm x_j \int \frac{\mathrm{d}\bm k}{2(2\pi)^3|\bm k|} \nonumber\\[1ex]
&\qquad\qquad\qquad\qquad \times e^{i|\bm k|(t_i-t_j)-\bm k\cdot (\bm x_i-\bm x_j)+ i(p\Omega_d t_i+q \Omega_{d'}t_j)-\frac{(\bm x_i-\bar{\bm x}_{l_d})^2}{\sigma^2_{l_d}}-\frac{(t_i-\bar{t}_d)^2}{T_d^2}-\frac{(\bm x_j-\bar{\bm x}_{m_{d'}})^2}{\sigma^2_{m_{d'}}}-\frac{(t_j-\bar{t}_{d'})^2}{T_{d'}^2}} \nonumber \\[1ex]
&= \sum_{l_d m_{d'}} \frac{\tilde{c}_{l_d}\tilde{c}_{m_{d'}}}{16\pi^3||\tilde{c}_d||||\tilde{c}_{d'}||}\int^\infty_{-\infty}\mathrm{d}t_i \int^\infty_{-\infty} \mathrm{d}t_j \Theta(t_j-t_i)\int \frac{\mathrm{d}\bm k}{|\bm k|} \nonumber\\[1ex]
&\qquad\qquad\qquad\qquad \times e^{i|\bm k|(t_i-t_j)-i\bm k \cdot(\bar{\bm x}_d-\bar{\bm x}_{d'}) + i(p\Omega_d t_i+q \Omega_{d'}t_j)-\frac{|\bm k|^2}{4}\beta_{l_dm_{d'}}-\frac{(t_i-\bar{t}_d)^2}{T_d^2}-\frac{(t_j-\bar{t}_{d'})^2}{T_{d'}^2}} \nonumber \\[1ex]
&= \sum_{l_d m_{d'}} \frac{\tilde{c}_{l_d}\tilde{c}_{m_{d'}}}{8iL_{l_dm_{d'}}||\tilde{c}_d||||\tilde{c}_{d'}||}\int^\infty_{-\infty}\mathrm{d}t_i \int^\infty_{-\infty} \mathrm{d}t_j \Theta(t_j-t_i)\int^\infty_0\mathrm{d}|\bm k| \left(e^{-i|\bm k|L_{l_dm_{d'}}}-e^{i|\bm k|L_{l_dm_{d'}}}\right)\nonumber \\[1ex]
&\qquad\qquad\qquad\qquad\qquad\qquad\qquad\qquad \times e^{i|\bm k|(t_i-t_j) + i(p\Omega_d t_i+q\Omega_{d'}t_j)-\frac{|\bm k|^2}{4}\beta_{l_dm_{d'}}-\frac{(t_i-\bar{t}_d)^2}{T_d^2}-\frac{(t_j-\bar{t}_{d'})^2}{T_{d'}^2}}
\end{align}
Now focus on the time integrals again and make the substitutions $t_i-t_j=x, t_i+t_j=y$, then
\begin{align}
&\frac{1}{2}\int^\infty_{-\infty}\mathrm{d}y \int^\infty_{-\infty} \mathrm{d}x \Theta(-x)e^{i|\bm k|x+ \frac{i}{2}(p (x+y)\Omega_d+q (y-x)\Omega_{d'}) -\frac{1}{T_d^2}((x+y)/2-\bar{t}_d)^2-\frac{1}{T_{d'}^2}((y-x)/2-\bar{t}_{d'})^2} \nonumber \\[1ex]
&= \frac{1}{2}\int^\infty_{-\infty}\mathrm{d}y \int^0_{-\infty} \mathrm{d}x e^{i|\bm k|x+ \frac{i}{2}(p(x+y)\Omega_d+q (y-x)\Omega_{d'}) -\frac{1}{T_d^2}((x+y)/2-\bar{t}_d)^2-\frac{1}{T_{d'}^2}((y-x)/2-\bar{t}_{d'})^2} \nonumber \\[1ex]
&= \frac{\pi T_d^2T_{d'}^2}{\sqrt{2(T_d^4+T_{d'}^4)}}e^{\frac{T_d^2T_{d'}^2}{2(T_d^4+T_{d'}^4)}\left(-|\bm k|^2(T_d^2+T_{d'}^2)-|\bm k|(2T_d^2(p\Omega_d-q\Omega_{d'})-4i(\bar{t}_d-\bar{t}_{d'}))\right)}\nonumber\\[1ex]
&\qquad\quad\times e^{\frac{1}{4(T_d^4+T_{d'}^4)(T_d^2+T_{d'}^2)}\left(4(\bar{t}_d-\bar{t}_{d'})^2((T_d^2+T_{d'}^2)^2-T_d^2T_{d'}^2)-(3T_d^4+T_{d'}^4)T_d^2T_{d'}^2(p\Omega_d-q\Omega_{d'})^2-4(p\Omega_d-q\Omega_{d'})(T_{d'}^2\bar{t}_d(3T_d^4+T_{d'}^4)+T_d^2\bar{t}_{d'}(T_d^4-T_{d'}^4))\right)}\nonumber\\[1ex]
&\qquad\quad\times \left(1+\text{erf}\left(\frac{(2(\bar{t}_d-\bar{t}_{d'})-i|\bm k|(T_d^2+T_{d'}^2)-iT_d^2(p\Omega_d-q\Omega_{d'}))T_dT_{d'}}{\sqrt{2(T_d^4+T_{d'}^4)(T_d^2+T_{d'}^2)}}\right)\right) 
\end{align}
which leads to
\begin{align}
&\sum_{l_d m_{d'}} \frac{(T_dT_{d'})^2\tilde{c}_{l_d}\tilde{c}_{m_{d'}}}{8iL_{l_dm_{d'}}\pi||\tilde{c}_d||||\tilde{c}_{d'}||\sqrt{2(T_d^4+T_{d'}^4)}} \int^\infty_0 \mathrm{d}|\bm k| e^{\frac{T_d^2T_{d'}^2}{2(T_d^4+T_{d'}^4)}\left(-|\bm k|^2(T_d^2+T_{d'}^2)-|\bm k|(2T_d^2(p\Omega_d-q\Omega_{d'})-4i(\bar{t}_d-\bar{t}_{d'}))\right)}\nonumber\\[1ex]
&\times  e^{-\frac{|\bm k|^2}{4}\beta_{l_dm_{d'}}}(e^{-i|\bm k|L_{l_dm_{d'}}}-e^{i|\bm k|L_{l_dm_{d'}}})\left(1+\text{erf}\left(\frac{(2(\bar{t}_d-\bar{t}_{d'})-i|\bm k|(T_d^2+T_{d'}^2)-iT_d^2(p\Omega_d-q\Omega_{d'}))T_dT_{d'}}{\sqrt{2(T_d^4+T_{d'}^4)(T_d^2+T_{d'}^2)}}\right)\right) \nonumber\\[1ex]
&\times e^{\frac{1}{4(T_d^4+T_{d'}^4)(T_d^2+T_{d'}^2)}\left(4(\bar{t}_d-\bar{t}_{d'})^2((T_d^2+T_{d'}^2)^2-T_d^2T_{d'}^2)-(3T_d^4+T_{d'}^4)T_d^2T_{d'}^2(p\Omega_d-q\Omega_{d'})^2-4(p\Omega_d-q\Omega_{d'})(T_{d'}^2\bar{t}_d(3T_d^4+T_{d'}^4)+T_d^2\bar{t}_{d'}(T_d^4-T_{d'}^4))\right)}\,.
\end{align}

\subsection{Retarded and Advanced Propagators}

In order to calculate the smeared retarded propagator, the restrictions $\sigma_{l_d} = \sigma_{m_{d'}}$ and $\bar{\bm x}_{l_d} = \bar{\bm x}_{m_{d'}}$ need to be imposed, i.e. that all systems have the same spatial smearing profile and centres. The smeared retarded propagator, for the same system, is then given by
\begin{align}
&G_R(\Lambda^p_d, \Lambda^q_{d}) = \sum_{l_d m_{d}} \frac{c_{l_d}c_{m_{d}}}{\pi^3||\tilde{c}_d||^2}\int^\infty_{-\infty}\mathrm{d}t_i \int^\infty_{-\infty} \mathrm{d}t_j \int \mathrm{d}\bm x_i \int \mathrm{d}\bm x_j \nonumber \\[1ex]
&\qquad\qquad\qquad \times e^{i\Omega_d(pt_i+qt_j)-\frac{(\bm x_i-\bar{\bm x}_{l_d})^2}{\sigma^2_{l_d}}-\frac{(t_i-\bar{t}_d)^2}{T_d^2}-\frac{(\bm x_j-\bar{\bm x}_{m_d})^2}{\sigma^2_{m_d}}-\frac{(t_j-\bar{t}_d)^2}{T_d^2}} \left(-\frac{\delta(t_j-t_i+|\bm x_i-\bm x_j|)}{4\pi |\bm x_i-\bm x_j|}\right)\nonumber \\[1ex]
&= -\sum_{l_d m_{d}} \frac{c_{l_d}c_{m_{d}}}{4\pi^4||\tilde{c}_d||^2}\int^\infty_{-\infty}\mathrm{d}t_i  \int \mathrm{d}\bm x_i \int \mathrm{d}\bm x_j \nonumber \\[1ex]
&\qquad\qquad \times e^{i\Omega_d(pt_i+q(t_i-|\bm x_i-\bm x_j|))-\frac{(\bm x_i-\bar{\bm x}_{l_d})^2}{\sigma^2_{l_d}}-\frac{(t_i-\bar{t}_d)^2}{T_d^2}-\frac{(\bm x_j-\bar{\bm x}_{m_d})^2}{\sigma^2_{m_d}}-\frac{(t_i-|\bm x_i-\bm x_j|-\bar{t}_d)^2}{T_d^2}} \frac{1}{|\bm x_i-\bm x_j|}
\end{align}
and now make the variable substitution 
\begin{align}
\bm x_i = \frac{1}{2}(u+v)+\bar{\bm x}_{l_d} \ , \ \ \bm x_j = \frac{1}{2}(u-v)+\bar{\bm x}_{m_d}
\end{align}
yielding
\begin{align}
& -\sum_{l_d m_{d}} \frac{c_{l_d}c_{m_{d}}}{32\pi^4||\tilde{c}_d||^2}\int^\infty_{-\infty}\mathrm{d}t_i  \int \mathrm{d}u \int \mathrm{d}v \nonumber \\[1ex]
&\qquad\qquad \times e^{i\Omega_d(pt_i+q(t_i-|v|))-\frac{1}{2\sigma^2_{l_d}}(|u|^2+|v|^2)-\frac{(t_i-\bar{t}_d)^2}{T_d^2}-\frac{(t_i-|v|-\bar{t}_d)^2}{T_d^2}} \frac{1}{|v|}\nonumber\\[1ex]
&= -\sum_{l_d m_{d}} \frac{(4\pi)^2c_{l_d}c_{m_{d}}}{32\pi^4||\tilde{c}_d||^2}\int^\infty_{-\infty}\mathrm{d}t_i  \int^\infty_0 \mathrm{d}|u||u|^2 \int^\infty_0 \mathrm{d}|v||v| \nonumber \\[1ex]
&\qquad\qquad \times e^{-\frac{1}{2\sigma^2_{l_d}}|u|^2}e^{-iq\Omega_d|v|-\frac{1}{2\sigma^2_{l_d}}|v|^2-\frac{|v|^2+2|v|\bar{t}_d}{T_d^2}}e^{i\Omega_d(p+q)t_i-2\frac{(t_i-\bar{t}_d)^2}{T_d^2}+\frac{2t_i|v|}{T_d^2}}\nonumber\\[1ex]
&= -\sum_{l_d m_{d}} \frac{(4\pi)^2c_{l_d}c_{m_{d}}}{32\pi^4||\tilde{c}_d||^2}\frac{\sqrt{\pi}}{\sqrt{2}\sigma_{l_d}^{-3}}\int^\infty_{-\infty}\mathrm{d}t_i  \int^\infty_0 \mathrm{d}|v||v| \nonumber \\[1ex]
&\qquad\qquad \times e^{-iq\Omega_d|v|-\frac{1}{2\sigma^2_{l_d}}|v|^2-\frac{|v|^2+2|v|\bar{t}_d}{T_d^2}}e^{i\Omega_d(p+q)t_i-2\frac{(t_i-\bar{t}_d)^2}{T_d^2}+\frac{2t_i|v|}{T_d^2}}\nonumber\\[1ex]
&= -\sum_{l_d m_{d}} \frac{(4\pi)^2c_{l_d}c_{m_{d}}e^{-\frac{1}{8}(p+q)^2(\Omega_dT_d)^2+i(p+q)\Omega_d\bar{t}_d}}{32\pi^4||\tilde{c}_d||^2}\frac{T_d\pi}{2\sigma_{l_d}^{-3}}  \int^\infty_0 \mathrm{d}|v||v|e^{-(\frac{1}{2\sigma^2_{l_d}}+\frac{1}{2T_d^2})|v|^2+\frac{1}{2}i\Omega_d(p-q)|v|}\nonumber\\[1ex]
&= -\sum_{l_d m_{d}} \frac{T_d^4\tilde{c}_{l_d}\tilde{c}_{m_{d}}e^{-\frac{1}{8}(p+q)^2(\Omega_dT_d)^2+i(p+q)\Omega_d\bar{t}_d}}{2\pi||\tilde{c}_d||^2(2T_d^2+\beta_{l_dl_d})^{3/2}}\Bigg[\frac{\sqrt{4T_d^2+2\beta_{l_dl_d}}}{T_d\sqrt{\beta_{l_dl_d}}}\nonumber\\[1ex]
&\qquad\qquad\qquad\qquad+\sqrt{\pi}\frac{\Omega_d(p-q)}{2}e^{-\frac{(\Omega_dT_d(p-q))^2\beta_{l_dl_d}}{8(2T_d^2+\beta_{l_dl_d})}}\left(i-\text{erfi}\left(\frac{\Omega_dT_d(p-q)\sqrt{\beta_{l_dl_d}}}{2\sqrt{4T_d^2+2\beta_{l_dl_d}}}\right)\right)\Bigg]
\end{align}
For different systems, the smeared retarded propagator is
\begin{align}
&G_R(\Lambda^p_d, \Lambda^q_{d'}) = \sum_{l_d m_{d'}}  \frac{c_{l_d}c_{m_{d'}}}{\pi^3||\tilde{c}_d||||\tilde{c}_{d'}||}\int^\infty_{-\infty}\mathrm{d}t_i \int^\infty_{-\infty} \mathrm{d}t_j \int \mathrm{d}\bm x_i \int \mathrm{d}\bm x_j \nonumber\\[1ex]
&\qquad\qquad \times e^{ i(p\Omega_dt_i+q\Omega_{d'}t_j)-\frac{(\bm x_i-\bar{\bm x}_{l_d})^2}{\sigma^2_{l_d}}-\frac{(t_i-\bar{t}_d)^2}{T_d^2}-\frac{(\bm x_j-\bar{\bm x}_{m_{d'}})^2}{\sigma^2_{l_d}}-\frac{(t_j-\bar{t}_{d'})^2}{T_{d'}^2}}\left(-\frac{\delta(t_j-t_i+|\bm x_i-\bm x_j|)}{4\pi |\bm x_i-\bm x_j|}\right) \nonumber \\[1ex]
&= -\sum_{l_d m_{d'}}  \frac{c_{l_d}c_{m_{d'}}}{4\pi^4||\tilde{c}_d||||\tilde{c}_{d'}||}\int^\infty_{-\infty}\mathrm{d}t_i \int \mathrm{d}x_i \int \mathrm{d}x_j \nonumber\\[1ex]
&\qquad\qquad \times e^{ i(p\Omega_dt_i+q\Omega_{d'}(t_i - |\bm x_i-\bm x_j|))-\frac{(\bm x_i-\bar{\bm x}_{l_d})^2}{\sigma^2_{l_d}}-\frac{(t_i-\bar{t}_d)^2}{T_d^2}-\frac{(\bm x_j-\bar{\bm x}_{m_{d'}})^2}{\sigma^2_{l_d}}-\frac{(t_i - |\bm x_i-\bm x_j|-\bar{t}_{d'})^2}{T_{d'}^2}}\frac{1}{ |\bm x_i-\bm x_j|}
\end{align}
and again make the variable substitution 
\begin{align}
\bm x_i = \frac{1}{2}(u+v)+\bar{\bm x}_{l_d} \ ,\ \ \bm x_j = \frac{1}{2}(u-v)+\bar{\bm x}_{m_{d'}}
\end{align}
yielding
\begin{align}
& -\sum_{l_d m_{d'}}  \frac{c_{l_d}c_{m_{d'}}}{32\pi^4||\tilde{c}_d||||\tilde{c}_{d'}||}\int^\infty_{-\infty}\mathrm{d}t_i \int \mathrm{d}u \int \mathrm{d}v \nonumber\\[1ex]
&\qquad\qquad \times e^{ i(p\Omega_dt_i+q\Omega_{d'}(t_i - |v|))-\frac{1}{4\sigma^2_{l_d}}(u+v)^2-\frac{(t_i-\bar{t}_d)^2}{T_d^2}-\frac{1}{4\sigma^2_{l_d}}(u-v+2L_{l_dm_{d'}})^2-\frac{(t_i - |v|-\bar{t}_{d'})^2}{T_{d'}^2}}\frac{1}{ |v|}\nonumber\\[1ex]
&= -\sum_{l_d m_{d'}}  \frac{c_{l_d}c_{m_{d'}}}{8\pi^2||\tilde{c}_d||||\tilde{c}_{d'}||}\int^\infty_{-\infty}\mathrm{d}t_i \int^\infty_0 \mathrm{d}|u||u|^2\int^1_{-1}\mathrm{d}(\cos{(u_\theta)}) \int^\infty_0 \mathrm{d}|v||v|\int^1_{-1}\mathrm{d}(\cos{(v_\theta)})  \nonumber\\[1ex]
&\ \times e^{ it_i(p\Omega_d+q\Omega_{d'})-\frac{(t_i-\bar{t}_d)^2}{T_d^2}-\frac{(t_i-\bar{t}_{d'})^2-2t_i|v|}{T_{d'}^2}}e^{-iq\Omega_{d'}|v|-\frac{|v|^2+2|v|\bar{t}_{d'}}{T_{d'}^2}-\frac{1}{2\sigma^2_{l_d}}(|v|^2-2|v|L_{l_dm_{d'}}\cos{(v_\theta)}+2L_{l_dm_{d'}}^2)}e^{-\frac{1}{2\sigma^2_{l_d}}(|u|^2+2|u|L_{l_dm_{d'}}\cos{(u_\theta)})}\nonumber\\[1ex]
&= -\sum_{l_d m_{d'}}  \frac{4c_{l_d}c_{m_{d'}}}{8\pi^2(\sigma^{-2}_{l_d}L_{l_dm_{d'}})^2||\tilde{c}_d||||\tilde{c}_{d'}||}\int^\infty_{-\infty}\mathrm{d}t_i \int^\infty_0 \mathrm{d}|u||u|e^{-\frac{1}{2\sigma^2_{l_d}}|u|^2}\text{sinh}(\sigma^{-2}_{l_d}L_{l_dm_{d'}}|u|)  \nonumber\\[1ex]
&\ \times \int^\infty_0 \mathrm{d}|v|e^{ it_i(p\Omega_d+q\Omega_{d'})-\frac{(t_i-\bar{t}_d)^2}{T_d^2}-\frac{(t_i-\bar{t}_{d'})^2-2t_i|v|}{T_{d'}^2}}e^{-iq\Omega_{d'}|v|-\frac{|v|^2+2|v|\bar{t}_{d'}}{T_{d'}^2}-\frac{1}{2\sigma^2_{l_d}}(|v|^2+2L_{l_dm_{d'}}^2)}\text{sinh}(\sigma^{-2}_{l_d}L_{l_dm_{d'}}|v|)\nonumber\\[1ex]
&= -\sum_{l_d m_{d'}}  \frac{2c_{l_d}c_{m_{d'}}e^{-\frac{1}{2\sigma^2_{l_d}}L_{l_dm_{d'}}^2}}{8\pi^2(\sigma^{-2}_{l_d}L_{l_dm_{d'}})^2||\tilde{c}_d||||\tilde{c}_{d'}||}\frac{\sqrt{2\pi}L_{l_dm_{d'}}}{\sigma^{-1}_{l_d}}\int^\infty_{-\infty}\mathrm{d}t_i \int^\infty_0 \mathrm{d}|v|\nonumber\\[1ex]
&\ \times e^{ it_i(p\Omega_d+q\Omega_{d'})-\frac{(t_i-\bar{t}_d)^2}{T_d^2}-\frac{(t_i-\bar{t}_{d'})^2-2t_i|v|}{T_{d'}^2}}e^{-iq\Omega_{d'}|v|-\frac{|v|^2+2|v|\bar{t}_{d'}}{T_{d'}^2}-\frac{1}{2\sigma^2_{l_d}}|v|^2}\text{sinh}(\sigma^{-2}_{l_d}L_{l_dm_{d'}}|v|)\nonumber\\[1ex]
&= -\sum_{l_d m_{d'}}  \frac{\sqrt{2}T_dT_{d'}c_{l_d}c_{m_{d'}}e^{-\frac{1}{2\sigma^2_{l_d}}L_{l_dm_{d'}}^2}}{4\pi\sigma_{l_d}^{-5}L_{l_dm_{d'}}||\tilde{c}_d||||\tilde{c}_{d'}||\sqrt{T_d^2+T_{d'}^2}}e^{\frac{-(\bar{t}_d-\bar{t}_{d'})^2-T_d^2T_{d'}^2(p\Omega_d+q\Omega_{d'})^2+i(p\Omega_d+q\Omega_{d'})(\bar{t}_dT_{d'}^2+\bar{t}_{d'}T_{d}^2)}{T_d^2+T_{d'}^2}}\nonumber\\[1ex]
&\ \times \int^\infty_0 \mathrm{d}|v|e^{-iq\Omega_{d'}|v|-\frac{|v|^2+2|v|\bar{t}_{d'}}{T_{d'}^2}-\frac{1}{2\sigma^2_{l_d}}|v|^2+\frac{1}{T_d^2+T_{d'}^2}(i|v|T_d^2(p\Omega_d+q\Omega_{d'})+2|v|\bar{t}_d+\frac{T_d^2}{T_{d'}^2}(|v|^2+2|v|\bar{t}_{d'}))}\text{sinh}(\sigma^{-2}_{l_d}L_{l_dm_{d'}}|v|)\nonumber\\[1ex]
&= -\sum_{l_d m_{d'}}  \frac{\sqrt{\pi}T_dT_{d'}\tilde{c}_{l_d}\tilde{c}_{m_{d'}}e^{-\frac{1}{2\sigma^2_{l_d}}L_{l_dm_{d'}}^2}}{8\pi L_{l_dm_{d'}}||\tilde{c}_d||||\tilde{c}_{d'}||\sqrt{T_d^2+T_{d'}^2+\beta_{l_dl_d}}}e^{\frac{-4(\bar{t}_d-\bar{t}_{d'})^2-T_d^2T_{d'}^2(p\Omega_d+q\Omega_{d'})^2+4i(p\Omega_d+q\Omega_{d'})(\bar{t}_dT_{d'}^2+\bar{t}_{d'}T_{d}^2)}{4(T_d^2+T_{d'}^2)}}\nonumber\\[1ex]
&\times\Bigg[e^{\frac{\left(\sigma^{-2}_{l_d}L_{l_dm_{d'}}\left(T_d^2+T_{d'}^2\right)-2(\bar{t}_{d'}-\bar{t}_{d})+i\left(p\Omega_dT_d^2-q\Omega_{d'}T_{d'}^2\right)\right)^2}{\left(T_d^2+T_{d'}^2\right)\left(2\sigma^{-2}_{l_d}\left(T_d^2+T_{d'}^2\right)+4\right)}}\left(1+i\text{erfi}\left(\frac{-i\sigma^{-2}_{l_d}L_{l_dm_{d'}}\left(T_d^2+T_{d'}^2\right)+2i(\bar{t}_{d'}-\bar{t}_{d})+\left(p\Omega_dT_d^2-q\Omega_{d'}T_{d'}^2\right)}{\sqrt{\left(T_d^2+T_{d'}^2\right)\left(2\sigma^{-2}_{l_d}\left(T_d^2+T_{d'}^2\right)+4\right)}}\right)\right)\nonumber\\[1ex]
&-e^{\frac{\left(\sigma^{-2}_{l_d}L_{l_dm_{d'}}\left(T_d^2+T_{d'}^2\right)+2(\bar{t}_{d'}-\bar{t}_{d})-i\left(p\Omega_dT_d^2-q\Omega_{d'}T_{d'}^2\right)\right)^2}{\left(T_d^2+T_{d'}^2\right)\left(2\sigma^{-2}_{l_d}\left(T_d^2+T_{d'}^2\right)+4\right)}}\left(1+i\text{erfi}\left(\frac{i\sigma^{-2}_{l_d}L_{l_dm_{d'}}\left(T_d^2+T_{d'}^2\right)+2i(\bar{t}_{d'}-\bar{t}_{d})+\left(p\Omega_dT_d^2-q\Omega_{d'}T_{d'}^2\right)}{\sqrt{\left(T_d^2+T_{d'}^2\right)\left(2\sigma^{-2}_{l_d}\left(T_d^2+T_{d'}^2\right)+4\right)}}\right)\right)\Bigg]\nonumber\\[1ex]
&= -\sum_{l_d m_{d'}}  \frac{\sqrt{\pi}T_dT_{d'}\tilde{c}_{l_d}\tilde{c}_{m_{d'}}e^{-\frac{1}{4}((\Omega_dT_d)^2+(\Omega_{d'}T_{d'})^2)+i(p\Omega_d\bar{t}_d+q\Omega_{d'}\bar{t}_{d'})}}{8\pi L_{l_dm_{d'}}||\tilde{c}_d||||\tilde{c}_{d'}||\sqrt{T_d^2+T_{d'}^2+\beta_{l_dl_d}}}\Bigg[\nonumber\\[1ex]
&e^{\frac{(p \Omega_dT_d^2-q\Omega_{d'}T_{d'}^2+2i(\bar{t}_{d'}-\bar{t}_{d}+L_{l_dm_{d'}}))^2}{4(\beta_{l_dm_{d'}}+T_d^2+T_{d'}^2)}}\left(1+i\text{erfi}\left(\frac{-i\sigma^{-2}_{l_d}L_{l_dm_{d'}}\left(T_d^2+T_{d'}^2\right)+2i(\bar{t}_{d'}-\bar{t}_{d})+\left(p\Omega_dT_d^2-q\Omega_{d'}T_{d'}^2\right)}{\sqrt{\left(T_d^2+T_{d'}^2\right)\left(2\sigma^{-2}_{l_d}\left(T_d^2+T_{d'}^2\right)+4\right)}}\right)\right)\nonumber\\[1ex]
&-e^{\frac{(p \Omega_dT_d^2-q\Omega_{d'}T_{d'}^2+2i(\bar{t}_{d'}-\bar{t}_{d}-L_{l_dm_{d'}}))^2}{4(\beta_{l_dm_{d'}}+T_d^2+T_{d'}^2)}}\left(1+i\text{erfi}\left(\frac{i\sigma^{-2}_{l_d}L_{l_dm_{d'}}\left(T_d^2+T_{d'}^2\right)+2i(\bar{t}_{d'}-\bar{t}_{d})+\left(p\Omega_dT_d^2-q\Omega_{d'}T_{d'}^2\right)}{\sqrt{\left(T_d^2+T_{d'}^2\right)\left(2\sigma^{-2}_{l_d}\left(T_d^2+T_{d'}^2\right)+4\right)}}\right)\right)\Bigg]
\end{align}
The smeared advanced propagator can then be obtained either from the identity $G_A(f, g) = G_R(f, g) - E(f, g)$ or from the property $G_A(\mf x, \mf x') = G_R(\mf x', \mf x)$ \cite{Perche_2024}. If the latter is used, $G_A(\Lambda^p_d, \Lambda^q_{d})$ for the same system case is merely the result for $G_R$ but with an additional minus sign multiplying $(p-q)$:
\begin{align}
&-\sum_{l_d m_{d}} \frac{T_d^4\tilde{c}_{l_d}\tilde{c}_{m_{d}}e^{-\frac{1}{8}(p+q)^2(\Omega_dT_d)^2+i(p+q)\Omega_d\bar{t}_d}}{2\pi||\tilde{c}_d||^2(2T_d^2+\beta_{l_dl_d})^{3/2}}\Bigg[\frac{\sqrt{4T_d^2+2\beta_{l_dl_d}}}{T_d\sqrt{\beta_{l_dl_d}}}\nonumber\\[1ex]
&\qquad\qquad\qquad\qquad-\sqrt{\pi}\frac{\Omega_d(p-q)}{2}e^{-\frac{(\Omega_dT_d(p-q))^2\beta_{l_dl_d}}{8(2T_d^2+\beta_{l_dl_d})}}\left(i+\text{erfi}\left(\frac{\Omega_dT_d(p-q)\sqrt{\beta_{l_dl_d}}}{2\sqrt{4T_d^2+2\beta_{l_dl_d}}}\right)\right)\Bigg]
\end{align}
In the case of different systems, the result for $G_A(\Lambda^p_d, \Lambda^q_{d'})$ has an additional minus sign multiplying the last 2 terms in the arguments of the exponentials and imaginary error functions contained in the square brackets of $G_R$:
\begin{align}
& -\sum_{l_d m_{d'}}  \frac{T_dT_{d'}\tilde{c}_{l_d}\tilde{c}_{m_{d'}}e^{-\frac{1}{4}((\Omega_dT_d)^2+(\Omega_{d'}T_{d'})^2)+i(p\Omega_d\bar{t}_d+q\Omega_{d'}\bar{t}_{d'})}}{8\sqrt{\pi} L_{l_dm_{d'}}||\tilde{c}_d||||\tilde{c}_{d'}||\sqrt{T_d^2+T_{d'}^2+\beta_{l_dl_d}}}\Bigg[\nonumber\\[1ex]
&e^{\frac{(p \Omega_dT_d^2-q\Omega_{d'}T_{d'}^2+2i(\bar{t}_{d'}-\bar{t}_{d}-L_{l_dm_{d'}}))^2}{4(\beta_{l_dm_{d'}}+T_d^2+T_{d'}^2)}}\left(1+i\text{erfi}\left(\frac{-i\sigma^{-2}_{l_d}L_{l_dm_{d'}}\left(T_d^2+T_{d'}^2\right)-2i(\bar{t}_{d'}-\bar{t}_{d})-\left(p\Omega_dT_d^2-q\Omega_{d'}T_{d'}^2\right)}{\sqrt{\left(T_d^2+T_{d'}^2\right)\left(2\sigma^{-2}_{l_d}\left(T_d^2+T_{d'}^2\right)+4\right)}}\right)\right)\nonumber\\[1ex]
&-e^{\frac{(p \Omega_dT_d^2-q\Omega_{d'}T_{d'}^2+2i(\bar{t}_{d'}-\bar{t}_{d}+L_{l_dm_{d'}}))^2}{4(\beta_{l_dm_{d'}}+T_d^2+T_{d'}^2)}}\left(1+i\text{erfi}\left(\frac{i\sigma^{-2}_{l_d}L_{l_dm_{d'}}\left(T_d^2+T_{d'}^2\right)-2i(\bar{t}_{d'}-\bar{t}_{d})-\left(p\Omega_dT_d^2-q\Omega_{d'}T_{d'}^2\right)}{\sqrt{\left(T_d^2+T_{d'}^2\right)\left(2\sigma^{-2}_{l_d}\left(T_d^2+T_{d'}^2\right)+4\right)}}\right)\right)\Bigg]
\end{align}

\subsection{Symmetric Propagator and Feynman Propagator}

The smeared symmetric propagator is then 
\begin{align} \label{Equation:SymProp}
&\Delta(\Lambda^p_d, \Lambda^q_{d'}) = G_R(\Lambda^p_d, \Lambda^q_{d'})+G_A(\Lambda^p_d, \Lambda^q_{d'}) \nonumber\\[1ex]
&= -\sum_{l_d m_{d}} \frac{\delta_{dd'}T_d^4\tilde{c}_{l_d}\tilde{c}_{m_{d}}e^{-\frac{1}{8}(p+q)^2(\Omega_dT_d)^2+i(p+q)\Omega_d\bar{t}_d}}{\pi||\tilde{c}_d||^2(2T_d^2+\beta_{l_dl_d})^{3/2}}\Bigg[\frac{\sqrt{4T_d^2+2\beta_{l_dl_d}}}{T_d\sqrt{\beta_{l_dl_d}}}\nonumber\\[1ex]
&\qquad\qquad\qquad\qquad-\sqrt{\pi}\frac{\Omega_d(p-q)}{2}e^{-\frac{(\Omega_dT_d(p-q))^2\beta_{l_dl_d}}{8(2T_d^2+\beta_{l_dl_d})}}\text{erfi}\left(\frac{\Omega_dT_d(p-q)\sqrt{\beta_{l_dl_d}}}{2\sqrt{4T_d^2+2\beta_{l_dl_d}}}\right)\Bigg]\nonumber\\[1ex]
&+\sum_{l_d m_{d'}}  \frac{i(1-\delta_{dd'})T_dT_{d'}\tilde{c}_{l_d}\tilde{c}_{m_{d'}}e^{-\frac{1}{4}((\Omega_dT_d)^2+(\Omega_{d'}T_{d'})^2)+i(p\Omega_d\bar{t}_d+q\Omega_{d'}\bar{t}_{d'})}}{4\sqrt{\pi}L_{l_dm_{d'}}||\tilde{c}_d||||\tilde{c}_{d'}||\sqrt{T_d^2+T_{d'}^2+\beta_{l_dl_d}}}\Bigg[\nonumber\\[1ex]
&e^{\frac{(p \Omega_dT_d^2-q\Omega_{d'}T_{d'}^2+2i(\bar{t}_{d'}-\bar{t}_{d}+L_{l_dm_{d'}}))^2}{4(\beta_{l_dm_{d'}}+T_d^2+T_{d'}^2)}}\text{erfi}\left(\frac{-i\sigma^{-2}_{l_d}L_{l_dm_{d'}}\left(T_d^2+T_{d'}^2\right)+2i(\bar{t}_{d'}-\bar{t}_{d})+\left(p\Omega_dT_d^2-q\Omega_{d'}T_{d'}^2\right)}{\sqrt{\left(T_d^2+T_{d'}^2\right)\left(2\sigma^{-2}_{l_d}\left(T_d^2+T_{d'}^2\right)+4\right)}}\right)\nonumber\\[1ex]
&+e^{\frac{(p \Omega_dT_d^2-q\Omega_{d'}T_{d'}^2+2i(\bar{t}_{d'}-\bar{t}_{d}-L_{l_dm_{d'}}))^2}{4(\beta_{l_dm_{d'}}+T_d^2+T_{d'}^2)}}\text{erfi}\left(\frac{i\sigma^{-2}_{l_d}L_{l_dm_{d'}}\left(T_d^2+T_{d'}^2\right)+2i(\bar{t}_{d'}-\bar{t}_{d})+\left(p\Omega_dT_d^2-q\Omega_{d'}T_{d'}^2\right)}{\sqrt{\left(T_d^2+T_{d'}^2\right)\left(2\sigma^{-2}_{l_d}\left(T_d^2+T_{d'}^2\right)+4\right)}}\right)\Bigg]
\end{align}
and finally the smeared Feynman propagator is 
\begin{align}
&G_F(\Lambda^p_d, \Lambda^q_{d'}) = \frac{1}{2}H(\Lambda^p_d, \Lambda^q_{d'})+\frac{i}{2}\Delta(\Lambda^p_d, \Lambda^q_{d'}) \nonumber\\[1ex]
&= \sum_{l_d m_{d}} \frac{\delta_{dd'}T_d^2\tilde{c}_{l_d}\tilde{c}_{m_{d}}e^{-\frac{1}{2}(T_d\Omega_d)^2+i(p+q)\Omega_d\bar{t}_d}}{2\pi||\tilde{c}_d||^2(\beta_{l_dm_d}+2T_d^2)^{\frac{3}{2}}}\Bigg[\sqrt{\beta_{l_dm_d}+2T_d^2}\left(1-\frac{i\sqrt{2}T_d}{\sqrt{\beta_{l_dl_d}}}e^{\frac{1}{8}(\Omega_dT_d(p-q))^2}\right)\nonumber\\[1ex]
&+ \frac{\sqrt{\pi}}{2}T_d^2 \Omega_d(p-q)e^{\frac{((p-q)\Omega_dT_d^2)^2}{4(\beta_{l_dm_d}+2T_d^2)}}\left(\text{erf}\left(\frac{T_d^2\Omega_d(p-q)}{2\sqrt{\beta_{l_dm_d}+2T_d^2}}\right)+i\text{erfi}\left(\frac{\Omega_dT_d(p-q)\sqrt{\beta_{l_dl_d}}}{2\sqrt{4T_d^2+2\beta_{l_dl_d}}}\right)\right)\Bigg]\nonumber\\[1ex]
&+ \sum_{l_d m_{d'}} \frac{(1-\delta_{dd'})T_dT_{d'}\tilde{c}_{l_d}\tilde{c}_{m_{d'}}e^{-\frac{1}{4}((T_d\Omega_d)^2+(T_{d'}\Omega_{d'})^2)+ i(p\Omega_d \bar{t}_d+q\Omega_{d'}\bar{t}_{d'})}}{8L_{l_dm_{d'}}\sqrt{\pi}||\tilde{c}_d||||\tilde{c}_{d'}||\sqrt{\beta_{l_dm_{d'}}+T_d^2+T_{d'}^2}}\Bigg[e^{\frac{(p \Omega_dT_d^2-q\Omega_{d'}T_{d'}^2+2i(\bar{t}_{d'}-\bar{t}_{d}+L_{l_dm_{d'}}))^2}{4(\beta_{l_dm_{d'}}+T_d^2+T_{d'}^2)}}\nonumber\\[1ex]
&\times\left(\text{erfi}\left(\frac{2(\bar{t}_{d'}-\bar{t}_d+L_{l_dm_{d'}})-i(p \Omega_dT_d^2-q\Omega_{d'}T_{d'}^2)}{2\sqrt{\beta_{l_dm_{d'}}+T_d^2+T_{d'}^2}}\right)-\text{erfi}\left(\frac{-i\sigma^{-2}_{l_d}L_{l_dm_{d'}}\left(T_d^2+T_{d'}^2\right)+2i(\bar{t}_{d'}-\bar{t}_{d})+\left(p\Omega_dT_d^2-q\Omega_{d'}T_{d'}^2\right)}{\sqrt{\left(T_d^2+T_{d'}^2\right)\left(2\sigma^{-2}_{l_d}\left(T_d^2+T_{d'}^2\right)+4\right)}}\right)\right)\nonumber \\[1ex]
&-e^{\frac{(p \Omega_dT_d^2-q\Omega_{d'}T_{d'}^2+2i(\bar{t}_{d'}-\bar{t}_{d}-L_{l_dm_{d'}}))^2}{4(\beta_{l_dm_{d'}}+T_d^2+T_{d'}^2)}}\nonumber\\[1ex]
&\times\left(\text{erfi}\left(\frac{2(\bar{t}_{d'}-\bar{t}_d-L_{l_dm_{d'}})-i(p \Omega_dT_d^2-q\Omega_{d'}T_{d'}^2)}{2\sqrt{\beta_{l_dm_{d'}}+T_d^2+T_{d'}^2}}\right)+\text{erfi}\left(\frac{i\sigma^{-2}_{l_d}L_{l_dm_{d'}}\left(T_d^2+T_{d'}^2\right)+2i(\bar{t}_{d'}-\bar{t}_{d})+\left(p\Omega_dT_d^2-q\Omega_{d'}T_{d'}^2\right)}{\sqrt{\left(T_d^2+T_{d'}^2\right)\left(2\sigma^{-2}_{l_d}\left(T_d^2+T_{d'}^2\right)+4\right)}}\right)\right)\Bigg]
\end{align}

\subsection{Mana Expression}\label{Appendix:ManaExpression}

Following Reference \cite{Nyström_Pranzini_Keski-Vakkuri_2024}, the amount of magic a state $\hat{\rho}$ has can be quantified using the measure known as mana, which they show is given by the expression
\begin{align}
M(\hat{\rho}_D(t)) = \log{\left(1-\rho_{22}+\frac{1}{3}\left[\left|\rho_{22}+2\text{Re}(\rho_{13})\right|+\left|\rho_{22}-\text{Re}(\rho_{13})-\sqrt{3}\text{Im}(\rho_{13})\right|+\left|\rho_{22}-\text{Re}(\rho_{13})+\sqrt{3}\text{Im}(\rho_{13})\right|\right]\right)}
\end{align}
for a 3-dimensional system, where $\rho_{ij}$ are the components of the density matrix $\hat{\rho}$. Particularizing to the density operator used in this work and assuming $\bar{t}_d = 0$, the mana is expressed as
\begin{align}\label{Equation:ManaUDW}
M(\hat{\rho}_D(t)) &= \log{}\Bigg(1-2W^{+-}_{11}+\frac{2}{3}\Bigg[\left|W^{+-}_{11}-2\text{Re}((\bar{W}^{*})^{++}_{11})\right|+\left|W^{+-}_{11}+\text{Re}((\bar{W}^{*})^{++}_{11})+\sqrt{3}\text{Im}((\bar{W}^{*})^{++}_{11})\right|\nonumber\\
&\qquad\qquad\qquad+\left|W^{+-}_{11}+\text{Re}((\bar{W}^{*})^{++}_{11})-\sqrt{3}\text{Im}((\bar{W}^{*})^{++}_{11})\right|\Bigg]\Bigg)\nonumber \\
&= \log\Bigg(1+ \frac{ T_1^2\lambda_1^2e^{- \frac{T_1^2 \Omega_1^2}{2}} }{ \pi \sqrt{\left(2 T_1^2 + 2\sigma^2_1 \right)^3} }\Bigg(
\frac{4}{3}\sqrt{\pi} T_1^2 \Omega_1 \left(1 - \operatorname{erf}\left( \frac{T_1^2 \Omega_1}{\sqrt{2 T_1^2 + 2\sigma^2_1}} \right) \right)e^{ \frac{ T_1^4 \Omega_1^2}{2 T_1^2 + 2\sigma^2_1} } 
- \sqrt{2 T_1^2 + 2\sigma^2_1} \nonumber \\
&\qquad+ \frac{1}{3} \left|\sqrt{\pi} T_1^2 \Omega_1 
\left(1 - \operatorname{erf}\left( \frac{T_1^2 \Omega_1}{\sqrt{2 T_1^2 +2\sigma^2_1}} \right) \right)
e^{ \frac{ T_1^4 \Omega_1^2}{2 T_1^2 +2\sigma^2_1} } 
- \frac{1}{2}\left(3-\sqrt{3} \frac{T_1}{\sigma_1}\right)\sqrt{2 T_1^2 + 2\sigma^2_1}   \right| \nonumber\\
&\qquad+ \frac{1}{3} \left|\sqrt{\pi} T_1^2 \Omega_1 \left(1 - \operatorname{erf}\left( \frac{T_1^2 \Omega_1}{\sqrt{2 T_1^2 + 2\sigma^2_1}} \right) \right)
e^{ \frac{ T_1^4 \Omega_1^2}{2 T_1^2 + 2\sigma^2_1} } 
- \frac{1}{2}\left(3+\sqrt{3} \frac{T_1}{\sigma_1} \right)\sqrt{2 T_1^2 + 2\sigma^2_1}   \right| \Bigg)\Bigg)
\end{align}

\section{Modified Pentagram Contextuality Scenario for the Qutrit }\label{Appendix:QutritContextScenario}

In this section, the general problem of finding a contextuality scenario for a 3-dimensional system is first considered, followed by a particularization to the pentagram setup detailed in Section \ref{Section:Contextuality}. 

Using the equivalence noted in section 2.3.2 of reference ~\cite{amaral2018graph}, a set of rank-1 projectors $\{\hat{P}_i\}_{i=1}^{n}$ is needed such that the eigenvalues of the operators $\{\hat{B}_i|\hat{B}_i=I-2\hat{P}_i\}_{i=1}^n$ are all dichotomic (e.g. $\pm 1$) and where $[\hat{P}_i, \hat{P}_j] = 0$ for some $i,j$ but $[\hat{P}_i, \hat{P}_j]\neq 0$ for other $i,j$. Furthermore, there needs to exist at least one state $\hat{\rho}$ such that the inequality $\sum_i^n \text{Tr}(\hat{\rho}\hat{P}_i) \leq \gamma$ is violated, with $\gamma$ marking the boundary between contextual and non-contextual states (or in the language of graphs, the maximum independence number of the corresponding compatibility graph \cite{amaral2018graph}). 

Since a qutrit system is used in this work, consider the case of a 3-dimensional Hilbert space with basis $\{|1\rangle, |0\rangle, |-1\rangle\}$ and define the vectors 
\begin{align}
|v_i \rangle = \frac{1}{\sqrt{|p_i|^2+|q_i|^2}}\left(p_i\cos{(\theta_i)}|1\rangle + q_i|0\rangle + p_i\sin{(\theta_i)}|-1\rangle\right)
\end{align}
where $p_i, q_i \in \mathbb C$ are arbitrary constants, and which it can be clearly seen that they have the properties $\langle v_i|v_i \rangle = 1$ and $|v_i\rangle \langle v_i|^2 = |v_i\rangle \langle v_i|$; thus defining a set of projectors $\{\hat{P}_i | \hat{P}_i = |v_i\rangle \langle v_i|\}_{i=1}^n$. These projectors have commutator
\begin{align}
[\hat{P}_i, \hat{P}_j] &= a
\begin{pmatrix}
(a_1p_ip_j^*-a_1^*p_i^*p_j)\cos{(\theta_i)}\cos{(\theta_j)} & p_iq_j^*a_1\cos{(\theta_i)}-q_i^*p_ja_1^*\cos{(\theta_j)} & x_2(\theta_i, \theta_j) \\ a_1q_ip_j^*\cos{(\theta_j)}-p_i^*q_ja_1^*\cos{(\theta_i)} & a_1q_iq_j^*-a_1^*q_i^*q_j & q_ip_j^*a_1\sin{(\theta_j)}-p_i^*q_ja_1^*\sin{(\theta_i)} \\ x_1(\theta_i, \theta_j) & p_iq_j^*a_1\sin{(\theta_i)}-q_i^*p_ja_1^*\sin{(\theta_j)} & (a_1p_ip_j^*-a_1^*p_i^*p_j)\sin{(\theta_i)}\sin{(\theta_j)}
\end{pmatrix}
\end{align}
where $a = \left((|p_i|^2+|q_i|^2)(|p_j|^2+|q_j|^2)\right)^{-1}$, $a_1 = p_i^*p_j\cos{(\theta_i-\theta_j)}+q_i^*q_j$, $x_1(\theta_i, \theta_j) = a_1p_ip_j^*\cos{(\theta_j)}\sin{(\theta_i)}-a_1^*p_i^*p_j\cos{(\theta_i)}\sin{(\theta_j)}$, and $x_2(\theta_i, \theta_j) = a_1p_ip_j^*\cos{(\theta_i)}\sin{(\theta_j)}-a_1^*p_i^*p_j\cos{(\theta_j)}\sin{(\theta_i)}$

To make further progress, one needs to define a set of contexts as $i, j$ pairs such that the above is 0 for these pairs but non-zero for other $i,j$ pairs. Before doing this, it is desirable to impose that the state $|-1\rangle$ be the one satisfying 
\begin{align}
\sum_i^n |\langle -1| v_i\rangle|^2 = \sum_i^n \frac{|p_i|^2}{|p_i|^2+|q_i|^2}\sin^2{(\theta_i)} = \gamma
\end{align}
-- since the ground state is used in this work -- along with the constraint
\begin{align}
\sum_i^n \langle 1| v_i\rangle\langle v_i |-1\rangle = \sum_i^n\frac{|p_i|^2\sin{(2\theta_i)}}{2(|p_i|^2+|q_i|^2)} \leq 0\,.
\end{align}

The system of equations that needs to be solved is then
\begin{align}
0 &= p_i^*p_j\cos{(\theta_i-\theta_j)}+q_i^*q_j \nonumber\\[1ex]
&\text{or} \nonumber\\[1ex]
0 &= (q_i^*p_iq_jp_j^*-q_ip_i^*q_j^*p_j)\cos{(\theta_i)}\cos{(\theta_j)} \,, \nonumber \\[1ex]
0 &= p_iq_j^*(p_i^*p_j\cos{(\theta_i-\theta_j)}+q_i^*q_j)\cos{(\theta_i)}-q_i^*p_j(p_ip_j^*\cos{(\theta_i-\theta_j)}+q_iq_j^*)\cos{(\theta_j)} \,,\nonumber \\[1ex]
0 &= -\frac{1}{2}|p_ip_j|^2\sin{(2(\theta_i-\theta_j))} + q_i^*q_jp_ip_j^*\cos{(\theta_i)}\sin{(\theta_j)}-q_iq_j^*p_i^*p_j\cos{(\theta_j)}\sin{(\theta_i)} \,,\nonumber \\[1ex]
0 &= (q_i^*p_iq_jp_j^*-q_ip_i^*q_j^*p_j)\cos{(\theta_i-\theta_j)} \,,\nonumber \\[1ex]
0 &= q_ip_j^*(p_i^*p_j\cos{(\theta_i-\theta_j)}+q_i^*q_j)\sin{(\theta_j)}-p_i^*q_j(p_ip_j^*\cos{(\theta_i-\theta_j)}+q_iq_j^*)\sin{(\theta_i)} \,,\nonumber \\[1ex]
0 &= (q_i^*p_iq_jp_j^*-q_ip_i^*q_j^*p_j)\sin{(\theta_i)}\sin{(\theta_j)} \,,\nonumber \\[1ex]
&\text{and} \nonumber \\[1ex]
\gamma &=\sum_i^n \frac{|p_i|^2}{|p_i|^2+|q_i|^2}\sin^2{(\theta_i)} \,,\nonumber\\[1ex]
0 &\geq \sum^n_i\frac{|p_i|^2\sin{(2\theta_i)}}{2(|p_i|^2+|q_i|^2)}\,.
\end{align}
Solving the first equation results in
\begin{align}
\theta_i = \theta_j + \phi_{ij}\ , \ -\frac{q_i^*q_j}{p_i^*p_j} = \cos{(\phi_{ij})}\ ,
\end{align}
and then the last two equations determine all $\theta_i$ in terms of $q_i, p_i$ and some reference angle $\theta^{*}$.

To simplify the equations, take $q_i = r_i\cos{(\alpha_i)}$ and $p_i = r_i\sin{(\alpha_i)}$. The system then becomes
\begin{align}\label{Equation:GroundStateContextConds}
\theta_i &= \theta_j+\cos^{-1}{(-\cot{(\alpha_i)}\cot{(\alpha_j)})}\,,\nonumber\\[1ex]
\gamma &=\sum_i^n \sin^2{(\alpha_i)}\sin^2{(\theta_i)} \,,\nonumber\\[1ex]
0 &\geq \frac{1}{2}\sum^n_i\sin^2{(\alpha_i)}\sin{(2\theta_i)}\,,
\end{align}
with the projectors taking the form
\begin{align}
\hat{P}_i = \begin{pmatrix}
\sin^2{(\alpha_i)}\cos^2{(\theta_i)}& \sin{(\alpha_i)}\cos{(\alpha_i)}\cos{(\theta_i)} & \sin{(\theta_i)}\cos{(\theta_i)}\sin^2{(\alpha_i)} \\ \sin{(\alpha_i)}\cos{(\alpha_i)}\cos{(\theta_i)} & \cos^2{(\alpha_i)} & \sin{(\alpha_i)}\cos{(\alpha_i)}\sin{(\theta_i)} \\ \sin{(\theta_i)}\cos{(\theta_i)}\sin^2{(\alpha_i)} & \sin{(\alpha_i)}\cos{(\alpha_i)}\sin{(\theta_i)} & \sin^2{(\alpha_i)}\sin^2{(\theta_i)}
\end{pmatrix}
\end{align}

A family of solutions to the above set of equations can be found by imposing the relations from the compatibility graph given by the pentagram edge set $\{(0,2),(0,3),(1,3),(1,4),(2,4)\}$, which has maximum independence number $\gamma=2$. The equation in the first line of the above is a set of equations that can be manipulated to produce an expression for $\theta_i$ in terms of $\theta_0$ and the angles $\phi_{ij}$:
\begin{align}
\theta_i = \{\theta_0, \theta_0-\phi_{03}-\phi_{13}, \theta_0-\phi_{02}, \theta_0-\phi_{03}, \theta_0-\phi_{24}-\phi_{02}\}\,,
\end{align}
plus the condition $\phi_{03}+\phi_{13} =\phi_{02}+\phi_{24}-\phi_{14}$. Using trigonometric sum identities for the inverse cosine, the extra condition can be manipulated to solve for $\alpha_3$, eventually yielding:
\begin{align}
\alpha_3 = \operatorname{arccot}\left(
\frac{\tan(\alpha_0)\tan(\alpha_1)\sin(\phi_{24} + \phi_{02} - \phi_{14})}{\sqrt{\tan^2(\alpha_1) + \tan^2(\alpha_0) - 2\tan(\alpha_1)\tan(\alpha_0)\cos(\phi_{24} + \phi_{02} - \phi_{14})}}\right)
\end{align}
The operators corresponding to a solution set will yield different functional behaviours that can be seen in the LHS of the non-contextuality inequality Eq. \eqref{Equation:QutritContextIneq} and the contextual fraction difference. One possible solution to Eqs. \eqref{Equation:GroundStateContextConds}, where the LHS of the non-contextuality inequality Eq. \eqref{Equation:QutritContextIneq} is larger than 0 for $\Omega_d = 0$, is
\begin{align}
\theta_0 &= \frac{3\pi}{4} \\[1ex]
\alpha_i &= \{\frac{9\pi}{10}, \frac{29\pi}{10}, \frac{\pi}{2}, \alpha_3, 0.54722012035572493\pi\}\,.
\end{align}
Another possible solution, where the difference between the last two diagonal terms of the matrix contained in the non-contextuality inequality LHS Eq. \eqref{Equation:QutritContextIneq} is the same as the far off-diagonal term, is
\begin{align}
\theta_0 &= 0.71549033656902731395587677242763007306802049110217\pi \\[1ex]
\alpha_i &= \{\frac{9\pi}{10}, 2.83737665013\pi, \frac{\pi}{2}, \alpha_3, 0.54722012035572493\pi\}\,.
\end{align}
This solution corresponds to no harvesting at $\Omega_d = 0$. Yet another possible solution to Eqs. \eqref{Equation:GroundStateContextConds}, where the LHS of the non-contextuality inequality Eq. \eqref{Equation:QutritContextIneq} is smaller than 0 for $\Omega_d = 0$, is
\begin{align}
\theta_0 &= \frac{17\pi}{20} \\[1ex]
\alpha_i &= \{0.83999268322\pi, \frac{3\pi}{4}, \frac{\pi}{2}, \alpha_3, \frac{\pi}{2}\}\,.
\end{align}


\end{document}